\documentclass[11pt]{article}
\usepackage{jcapmod}

\usepackage{mathtools}
\usepackage{booktabs}
\usepackage[english]{babel}
\usepackage{amsmath,amssymb,amsbsy,amstext, amsthm, simplewick, amsfonts}
\usepackage{hyperref}
\usepackage{graphicx}
\usepackage[small]{caption}
\usepackage{siunitx}
\usepackage{upgreek}
\usepackage{framed}
\usepackage{wrapfig}
\usepackage{multirow}
\usepackage{bbm}
\usepackage[svgnames,dvipsnames,x11names]{xcolor}
\usepackage[utf8x]{inputenc}
\usepackage{selinput}
\usepackage{bm}
\usepackage{float}
\usepackage{geometry}
\usepackage{yfonts}
\usepackage{caption}
\usepackage{subcaption}
\usepackage{subfloat}
\usepackage{sidecap}
\usepackage{longtable}
\usepackage{physics}
\usepackage{anyfontsize}
\usepackage{dsfont}

\usepackage{colortbl}
\definecolor{summersky}{cmyk}{0.71,0.33,0,0.5}
\definecolor{flamingo}{cmyk}{0,0.51,0.71,0.5}
\definecolor{rp}{cmyk}{0.2, 1, 0.6, 0}
\definecolor{pacificblue}{cmyk}{0.95,0.3,0, 0.5}
\definecolor{gray60}{cmyk}{0.4,0.4,0,0.8}

\usepackage[framemethod=default]{mdframed}
\newmdenv[skipabove=7pt,
skipbelow=7pt,
rightline=false,
leftline=false,
topline=false,
bottomline=false,
backgroundcolor=pacificblue!10,
linecolor=gray,
innerleftmargin=5pt,
innerrightmargin=5pt,
innertopmargin=2pt,
innerbottommargin=10pt,
leftmargin=0cm,
rightmargin=0cm,
linewidth=4pt]{eBox}
\newmdenv[skipabove=7pt,
skipbelow=7pt,
rightline=false,
leftline=false,
topline=false,
bottomline=false,
backgroundcolor=gray!10,
linecolor=gray,
innerleftmargin=5pt,
innerrightmargin=5pt,
innertopmargin=-5pt,
innerbottommargin=5pt,
leftmargin=0cm,
rightmargin=0cm,
linewidth=4pt]{eBox2}

\definecolor{blue3}{RGB}{31, 119, 180}
\definecolor{red3}{RGB}{	214, 39, 40}
\definecolor{orange3}{RGB}{255, 127, 14}
\definecolor{green3}{RGB}{44, 160, 44}
\definecolor{repBlue}{RGB}{31, 119, 180}
\definecolor{repRed}{RGB}{	214, 39, 40}
\definecolor{repGreen}{RGB}{44, 160, 44}

\renewcommand{\(}{\left(}
\renewcommand{\)}{\right)}
\renewcommand{\[}{\left[}
\renewcommand{\]}{\right]}

\newcommand \nn {\nonumber}

\def\be{\begin{equation}}
\def\ee{\end{equation}}
\newcommand{\bea}{\begin{eqnarray}}
\newcommand{\eea}{\end{eqnarray}}
\def\fnl{f_{\rm NL}}

\newcommand{\rom}[1]{\uppercase\expandafter{\romannumeral #1\relax}}

\makeatletter

\newcommand{\Rmnum}[1]{\expandafter\@slowromancap\romannumeral #1@}
\makeatother

\usepackage{colortbl}
\definecolor{lightgreen}{cmyk}{0.2, 0, 0.2, 0.2}
\definecolor{lightgray}{cmyk}{0.1,0.2,0,0.1}
\definecolor{lightgray2}{cmyk}{0.1,0.1,0,0.1}

\makeatletter
\newlength{\apb@width}
\newcommand{\autoparbox}[2][c]{\settowidth{\apb@width}{#2}\parbox[#1]{\apb@width}{#2}}

\makeatother


\def\beq{\begin{equation}}
\def\eeq{\end{equation}}

\newcommand{\calR}{\mathcal{R}}

\allowdisplaybreaks[1]
\rightline{YITP-22-48}
\title{Highly non-Gaussian tails and primordial black holes from single-field inflation
}

\author[a,b]{Yi-Fu Cai,}
\author[a,b]{Xiao-Han Ma,}
\author[c,d,e]{Misao Sasaki,}
\author[f]{Dong-Gang Wang,}
\author[a,g]{Zihan Zhou}

\affiliation[a]{Deep Space Exploration Laboratory/School of Physcial Sciences, \\
University of Science and Technology of China, Hefei, Anhui 230026, China}
\affiliation[b]{CAS Key Laboratory for Researches in Galaxies and Cosmology/Department of Astronomy, School of Astronomy and Space Science, \\University of Science and Technology of China, Hefei, Anhui 230026, China}
\affiliation[c]{Kavli Institute for the Physics and Mathematics of the Universe (WPI), UTIAS,\\ The University of Tokyo, Chiba 277-8583, Japan}
\affiliation[d]{Center for Gravitational Physics and Quantum Information, Yukawa Institute for Theoretical Physics,\\ Kyoto University, Kyoto 606-8502, Japan}
\affiliation[e]{Leung Center for Cosmology and Particle Astrophysics, National Taiwan University, Taipei 10617}
\affiliation[f]{Department of Applied Mathematics and Theoretical Physics,\\ University of Cambridge, Wilberforce Road, Cambridge, CB3 0WA, UK}
\affiliation[g]{Department of Physics, Princeton University, Princeton, NJ 08544, USA}

\emailAdd{yifucai@ustc.edu.cn}
\emailAdd{mxh171554@mail.ustc.edu.cn}
\emailAdd{misao.sasaki@ipmu.jp}
\emailAdd{dgw36@cam.ac.uk}
\emailAdd{zihanz@princeton.edu}

\date{}

\abstract{
For primordial perturbations, deviations from Gaussian statistics on the tail of the probability distribution can be associated with non-perturbative effects of inflation. In this paper, we present some particular examples in which the tail of the distribution becomes highly non-Gaussian although the statistics remains almost Gaussian in the perturbative regime. We begin with an extension of the ultra-slow-roll inflation that incorporates a transition process, where the inflaton climbs up a tiny potential step at the end of the non-attractor stage before it converges to the slow-roll attractor. Through this example, we identify the key role of the off-attractor behaviour for the upward-step transition, and then extend the analysis to another type of the transition with two slow-roll stages connected by a tiny step. We perform both the perturbative and non-perturbative analyses of primordial fluctuations generated around the step in detail, and show that the tiny but nontrivial transition may affect large perturbations in the tail of the distribution, while the perturbative non-Gaussianity remains small. Our result indicates that the non-Gaussian tails can have rich phenomenology which has been overlooked in conventional analyses. We also study the implications of this non-Gaussian tail for the formation of primordial black holes, and find that their mass fraction can be parametrically amplified by several orders of magnitudes in comparison with the case of the Gaussian distribution. Additionally, we also discuss a mechanism of primordial black holes formation for this upward step inflation model by trapping the inflaton in the bottom of the step.
}

\begin{document}

\maketitle

\section{Introduction}

The latest cosmological observations suggest that the primordial curvature perturbation $\calR$ is nearly Gaussian \cite{Planck:2018jri, Planck:2019kim}, which is in agreement with the predictions of the single-field slow-roll inflation \cite{Maldacena:2002vr}. For future cosmological surveys, any detection of deviations from the Gaussian statistics will reveal important information about the primordial universe. Therefore, primordial non-Gaussianity is one of the major targets in upcoming experiments \cite{Meerburg:2019qqi, Achucarro:2022qrl}. So far theoretical studies on primordial non-Gaussianity have been mostly performed in the framework of cosmological perturbation theory, where cosmological correlators, such as the bispectrum and trispectrum, are examined by using perturbative approaches (see \cite{Chen:2010xka} for a review). These correlation functions provide a good description of the non-Gaussian statistics when the perturbation is small.

Except for the $n$-point correlators, there in principle exist significant phenomenological implications in the probability distribution of curvature perturbations that are not captured by perturbative approaches (see \cite{Chen:2018uul, Chen:2018brw, Panagopoulos:2019ail, Panagopoulos:2020sxp, Celoria:2021vjw, Cohen:2021jbo, Ezquiaga:2019ftu, Figueroa:2020jkf, Pattison:2021oen, Achucarro:2021pdh, Ahmadi:2022lsm} for recent discussions). In particular, the perturbation theory breaks down at the tail of the distribution where fluctuations are large and rare. Therefore, if deviations from Gaussianity appear in the tail, cosmological correlators are no longer appropriate in describing the physics involving large and rare fluctuations. The nontrivial behaviour in the tail of the distribution, dubbed {\it the non-Gaussian tail}, has been actively discussed in connection with primordial black holes (PBHs). It is speculated that black holes may have formed in the early Universe due to the presence of large curvature fluctuations \cite{Carr:1974nx} and these objects could serve an inspiring tool to probe the unknown physics in the
very early universe \cite{Khlopov:2008qy, Sasaki:2018dmp, Carr:2016drx}. 
Since the large fluctuations are highly sensitive to the tail of the probability distribution function, non-Gaussianities in the initial condition from inflation may play a crucial role in determining the abundance of PBHs \cite{Franciolini:2018vbk, Biagetti:2018pjj, Atal:2018neu, Passaglia:2018ixg, Atal:2019cdz, Atal:2019erb, Meng:2022ixx,Taoso:2021uvl, Biagetti:2021eep, Davies:2021loj, Hooshangi:2022lao, DeLuca:2022rfz}.

In most of previous studies, both perturbative and non-perturbative regimes are governed by the same physics, and thus the non-Gaussian tails always have an exponential form. For instance, ultra-slow-roll (USR) inflation is one of the well-studied models in which the inflaton undergoes a period when the slow-roll (SR) conditions are violated and an $\mathcal{O}(1)$ local non-Gaussianity is generated \cite{Namjoo:2012aa, Chen:2013aj, Chen:2013eea, Cai:2017bxr}.
In the simplest model of USR inflation, it was shown that the resulting non-Gaussian tail is determined by the amplitude of the bispectrum represented by the parameter $\fnl$, which is defined in the perturbative analysis.
%
%
In principle, however, we remind that one should go beyond perturbation theory to properly understand the behaviour of the tail of distribution. 
For example, if there are some non-perturbative effects that affect only large fluctuations, it is very likely that one gets a large non-Gaussian tail even if the perturbative non-Gaussian parameters like $\fnl$, $g_{\rm nl}$, $\tau_{\rm nl}$, etc. are small. In other words, the behaviour of the distribution function at $|\calR|\ll 1$ and that at $|\calR|\sim 1$ can be uncorrelated. Interestingly, this gives rise to a hypothetical possibility that the distribution may be perfectly Gaussian at $|\calR|$ is small, while it could have a highly non-Gaussian bump at the tail, as sketched in figure \ref{fig:pdf1}:

\begin{figure}[htb]
  \centering
  \includegraphics[scale = 0.6]{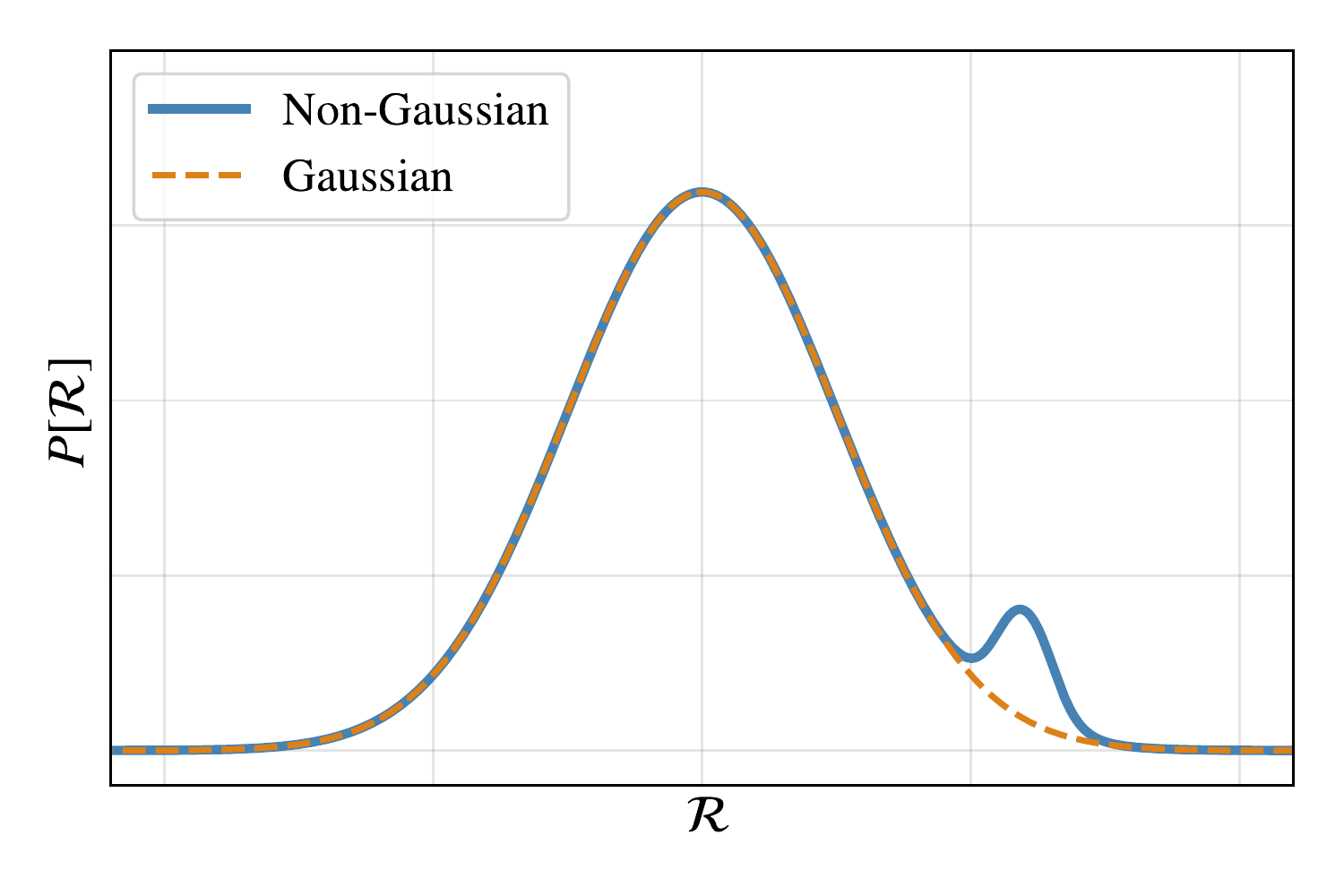}
  \caption{A speculative probability distribution of the curvature perturbation $\calR$, where the dotted orange curve is the Gaussian distribution which fits well at small $\calR$ but not at the tail.}
    \label{fig:pdf1}
\end{figure}

Recently we proposed a specific realization of the above phenomenon \cite{Cai:2021zsp}, where an upward step in the inflaton potential generates a highly non-Gaussian tail while the perturbative non-Gaussianity remains small.
In this paper, we extend the study upon the aforementioned interesting phenomenon into more general examples by performing much comprehensive and detailed analyses. In particular, we consider the canonical single-field inflation with a tiny upward step in the potential and investigate both the USR-SR and SR-SR transitions. In both cases, by deriving the exact background solutions, we identify the crucial role of {\it off-attractor phase trajectories} around the upward-step transition. Afterwards, we perform the analysis of perturbations via the $\delta N$ formalism, which can clearly reveal the essential role of the off-attractor trajectories around the step with a quantitative description. To be concrete, they affect small and large fluctuations differently. The small fluctuations are still in the perturbative regime, which can be analyzed using the conventional approach for the power spectrum, the bispectrum, etc. However, the large fluctuations are {\it non-perturbatively} affected by the step. That is to say, the inflaton field may not be able to move forward if the field fluctuation is so large that it makes the forward momentum too small to climb the step. This effect gives rise to the non-perturbative expression of the curvature perturbation, which consequently leads to nontrivial shape at the tail of the probability distribution. As an application, we also estimate how this novel non-Gaussian tail affects the abundance of PBHs. 

We mention that we exclusively rely on the $\delta N$ formalism in this paper, while the stochastic $\delta N$ formalism~\citep{Vennin:2015hra} 
is a powerful tool for analyzing non-perturbative features of inflatinary perturbations, particularly when the quantum diffusion effects are 
significant~\citep{Pattison:2017mbe,Animali:2022otk,Pattison:2019hef,Jackson:2022unc,Tada:2021zzj,Vennin:2020kng,Pattison:2021oen,Firouzjahi:2018vet}. 
In this paper, our interest is mainly in the case when the quantum diffusion effects are small. However, 
depending on the choice of the model parameters, it may play an important role. This point is further discussed in Section \ref{NG tail}.

The paper is organized as follows. In Section 2, we discuss the background dynamics of two models of upward-step transition in canonical single-field inflation, with a focus on the off-attractor trajectories.
Then we present the detailed analyses of perturbations by using the $\delta N$ formalism. Section 3 comprises the major part of this paper, in which we discuss the nontrivial tail behaviour in the probability distribution of the curvature perturbation. Applying the results of Section 2 in a concrete example, we demonstrate how a non-perturbative effect during inflation (a tiny upward step here) can lead to a significant modification in the non-Gaussian tail. In Section 4, we consider the corresponding implications into the formation of PBHs. We analyse the curvature perturbation in a much realistic model with an inflection-point potential, and compute the mass fraction of PBHs in the presence of the non-perturbative non-Gaussian tail. We also briefly discuss another mechanism for the PBH formation, i.e. the formation of PBHs due to trapping of the inflaton field at the local minimum of the potential at the step. We conclude in Section 5 with an outlook of future directions. 
Throughout the whole paper, we use the natural units $c = \hbar = 1$, and the reduced Planck mass $M_{\rm pl}^2 = 1/8\pi G$.

\section{Upward-step transitions}
\label{sec:ustep}

In this section we investigate the background evolution of canonical single-field inflation with an upward step in the potential, and then derive the $\delta N$ formula based on the background solutions. 
First, we perform a detailed analysis of a simple model of USR-SR upward-step transition. We adopt non-attractor initial conditions and evolve them through the step to the slow-roll phase after the transition.
With this concrete example, we illustrate the nontrivial effect of the off-attractor phase-space trajectories around the step. 
After that, we extend the analysis into an SR-SR upward-step transition model where non-attractor trajectories also play a crucial role because of the quantum fluctuations, even though the fiducial, classical trajectory is a slow-roll one before the step.

\subsection{USR-SR transition}\label{sec:USR_SR}

\subsubsection{A recap of USR inflation}

Let us begin with a brief review of the simplest USR inflation, where the inflaton rolls on a constant potential $V(\phi)=V_0$,
\be
 \ddot\phi+3H\dot\phi=0\,.
\ee
Unlike the conventional slow-roll inflation in which the inflaton is in an attractor phase with $\dot\phi$ being determined by the value of $\phi$, there is no attractor phase in this case and hence $\dot\phi$ remains to be a dynamical degree of freedom, independent of $\phi$.

Using the number of e-folds $n$ as the time variable via $dn = Hdt$, the background equations can be approximately written as
\be
 \frac{d^2\phi}{d n^2} + 3\frac{d\phi}{d n} = 0 ~, ~~~~~~ 3H^2\simeq V_0 ~.
\ee
We denote the initial values at $n=n_i$ by $\phi(n_i)$ and $\pi(n_i)=d\phi/dn(n_i)$.
Then, we have
\begin{align}
    \phi(n) &= \phi(n_i) + \frac{\pi(n_i)}{3}(1-e^{-3(n-n_i)}) ~, \label{phi solution 1 in variable N}\\
    \pi(n) &=\dv{\phi}{n_i} = \pi(n_i) e^{-3(n-n_i)} ~. \label{pi solution 1 in variable N}
\end{align}
Note that we have the following relation,
\be \label{pic}
\pi(n)+3\phi(n)= \pi(n_i) + 3\phi(n_i).
\ee
Fixing the final values as $\phi(n_c)=\phi_c$ and $\pi(n_c)=\pi_c$, the number of e-folds $N_{USR}\equiv n_c-n_i$
may be expressed as a function of the initial values at $n=n_i$, which we denote by $\phi_i$ and $\pi_i$,
\be \label{USR-N}
N_{USR}(\phi_i, \pi_i; \phi_c,\pi_c) =\frac{1}{3} \log\[\frac{\pi_i}{\pi_c}\]= \frac{1}{3} \log\[ \frac{\pi_i}{\pi_i+3(\phi_i-\phi_c)}\].
\ee
In the simplest case, the USR stage is assumed to end at $\phi=\phi_c$ independent of $\pi_c$ which is expected to be extremely small as $\pi$ has decayed exponentially. Therefore, when we take the variation of $N_{USR}$ to evaluate the curvature perturbation, we only fix $\phi_c$.
As a result, the expansion history of the Universe depends on the initial conditions of both $\phi_i$ and $\pi_i$, which differs from the single-clock behaviour of slow-roll inflation where $\phi$ is the only independent degree of freedom. Because of this, the USR inflation generates $\mathcal{O}(1)$ local non-Gaussianity with $\fnl =5/2$, and this simple model has been known as an example of canonical single-field models that violates Maldacena's consistency relation \cite{Namjoo:2012aa}\footnote{The same dominant mode of primordial perturbations was obtained as well in the single-field matter bounce cosmology \cite{Cai:2009fn, Quintin:2015rta, Li:2016xjb}.}.

Meanwhile, it has been well recognized that, this USR evolution is allowed only in a limited period of time, and for a complete description of inflation one still needs a subsequent slow-roll (SR) phase \cite{Cai:2016ngx, Cai:2017bxr}. Therefore, it is important to take into account the {\it USR-SR transition} in the realistic model building. The consequences of both smooth and sharp transitions have been investigated in detail in \cite{Cai:2017bxr}, which established that the final size of local non-Gaussianity is very sensitive to the transition process. When the transition is smooth, the USR result  $\fnl =5/2$ will be completely erased in the subsequent evolution; while it remains non-vanishing for sharp transitions with the maximum value of local $\fnl$ being $5/2$. These results were further confirmed by the derivation of a generalized consistency relation using the background field method \cite{Suyama:2021adn}.

\begin{figure}[htb]
    \centering
    \includegraphics[scale = 0.3]{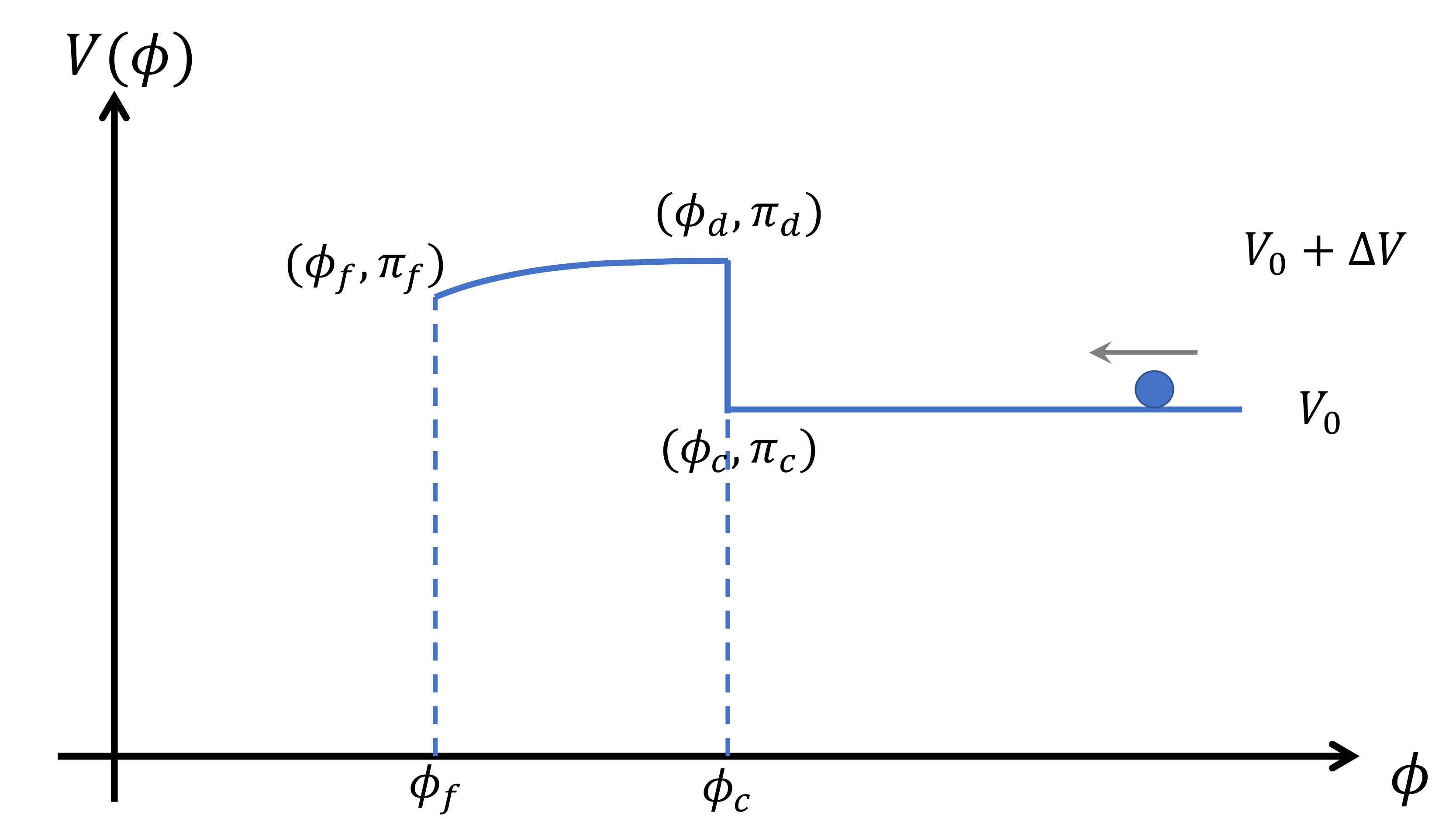}
    \caption{A sketch plot of the potential for an upward step transition. }
    \label{fig:step}
\end{figure}

\subsubsection{USR-SR transition with an upward step}
\label{sec:usr-sr}

Now we consider a USR-SR transition with an upward step at $\phi=\phi_c$, whose potential is illustrated in figure \ref{fig:step}. Initially, the inflaton is in the USR phase at $\phi>\phi_c$, moving toward the negative $\phi$ direction. The USR phase ends at $\phi=\phi_c$, and the inflaton climbs up the upward step $\Delta V$ at a cost of spending some of its kinetic energy $\pi_c$, assuming that $\pi_c$ is large enough to allow the upward jump. After the step, there is a short relaxation stage of a non-attractor phase, and the inflaton eventually reaches the slow-roll attractor after the relaxation stage. For simplicity we assume that inflation ends at $\phi=\phi_f$.

\paragraph{Background dynamics}
With the above picture in mind, let us derive the background solution of $\phi(t)$ from the USR stage to the final SR stage. 

At $\phi_c$, the inflaton velocity drops instantly in order to climb up the step. According to energy conservation the velocity right after the step is determined as
\be \label{picpid}
\pi_d = - \sqrt{\pi_c^2-6\frac{\Delta V}{V}} ~,
\ee
which serves as the initial condition for the subsequent stage. The minus sign comes from the assumption that the inflaton evolves toward the negative $\phi$ direction. For later convenience, we define the ratio of these two velocities as follows,
\be \label{g}
g\equiv \frac{\pi_d}{\pi_c}\quad (0< g <1) ~.
\ee
This will be the key parameter in our analysis below. When $\Delta V \rightarrow 0$, we have $g \rightarrow 1$, and the system goes back to the cases discussed in Ref.~\cite{Cai:2017bxr}. For $g\ll 1$, the effects of the step at the transition become significant.

Afterwards, the inflaton spends a few e-folds in a non-attractor, relaxation phase before it enters the slow-roll attractor stage. For the purpose of describing this transition process, we focus on small inflaton displacements, and parameterise the potential as
\begin{align}
    V(\phi) = (V_0 + \Delta V)\[1+  \sqrt{2\epsilon_V}(\phi -\phi_c)
    + \frac{1}{2}\eta_V(\phi -\phi_c)^2 \] ; \quad \phi<\phi_c \,.
    \label{eq:potential}
\end{align}
where $\epsilon_V$ and $\eta_V$ are the slow-roll parameters defined at $\phi_c$. For simplicity, we assume the above form is valid until the inflaton reaches $\phi_f$. Then, the equation of motion for $\phi$ is given by
\begin{align}
    \frac{d^{2} \phi}{d n^{2}}+3 \frac{d \phi}{dn}+3
    \sqrt{2 \epsilon_{V}}+3 \eta_{V}\left(\phi-\phi_{c}\right) = 0 ~,
    \label{background equation in slow-roll}
\end{align}
where we have approximated the Hubble parameter to be a constant given by $3H^2=V_0+\Delta V$.

Setting $n=n_c$ at $\phi=\phi_c$ and $\pi=\pi_d$, we obtain the analytical solution to be
\begin{align}
 \phi (n) &=\frac{s-3-h}{s(s-3)} \pi_{d} e^{\frac{1}{2}(s-3) (n-n_c)}-\frac{s+3+h}{s(s+3)} \pi_{d}
    e^{-\frac{1}{2}(s+3) (n-n_c)}+\frac{2 \pi_{d} h}{s^{2}-9}+\phi_{c} ~.\label{phi solution of slow-roll}\\
 \pi (n) &=\frac{d\phi(n)}{dn}=\frac{s-3-h}{2s} \pi_{d} e^{\frac{1}{2}(s-3) (n-n_c)}+\frac{s+3+h}{2s} \pi_{d}
    e^{-\frac{1}{2}(s+3) (n-n_c)}  ~.\label{pi solution of slow-roll}
\end{align}
where we have defined
\be \label{hs}
 h\equiv 6\sqrt{2\epsilon_V}/\pi_d , ~~~~~~~~ s \equiv 3\sqrt{1-4\eta_V/3} \simeq 3-2\eta_V .
\ee
Note that, $h$ is negative and can be written in terms of the ratio of two field velocities $h=-6\pi_f/\pi_d$, where $\pi_f\simeq-\sqrt{2\epsilon_V}$ is 
the velocity at $\phi=\phi_f$. The $h$ parameter can be any negative real number. For $h=-6$, we have $\pi_f=\pi_d$ and thus the inflaton evolution joins the slow-roll attractor immediately after the step. For other values of $h$, a relaxation phase is expected before the inflaton reaches the slow-roll attractor. From the solution given by \eqref{phi solution of slow-roll} and \eqref{pi solution of slow-roll}, we observe that there exists an equality,
\be
 \pi(n) + \frac{s+3}{2} \phi(n)
 = \pi_d \left[\(1-\frac{h}{s-3}\) e^{\frac{1}{2}(s-3) (n-n_c)}  +\frac{h}{s-3}\right]+\frac{s+3}{2} \phi_c\,,
\ee
which is an analog of the relation \eqref{pic} for the USR inflation.

\paragraph{Number of e-folds}
With the exact background solution, it is possible to obtain an expression for the number of e-folds, which plays the central role in the perturbation analysis in Subsection \ref{sec:deltaN}. Let $n_f$ be the number of e-folds at $\phi_f$, $\phi(n_f)=\phi_f$. We consider $n_f-n_c\gg1$ so that the second term in \eqref{phi solution of slow-roll} can be neglected.
Accordingly, we have
\begin{align}
    \phi_f \simeq \frac{s-3-h}{s(s-3)} \pi_{d} e^{\frac{1}{2}(s-3) (n_f-n_c)} +
    \frac{2 \pi_{d} h}{s^{2}-9}+\phi_{c} ~.
\label{phi_e approximate solution}
\end{align}
This gives the number of e-folds after the step $N_{SR}\equiv n_f-n_c$ as a function of $\pi_d$,
\begin{align}
    N_{SR}(\pi_d; \phi_f) &= \frac{2}{s-3} \log \left\{ \frac{s(s-3)}{s-3-h}\left[ \frac{(\phi_f - \phi_c)}{\pi_d} - \frac{2h}{s^2 - 9}\right] \right\} \nn\\
    &\simeq \frac{1}{\eta_V} \log\( -2\eta_V \pi_d-6\sqrt{2\epsilon_V} \) + {\rm const.} \,,
    \label{SR-N}
\end{align}
where we have separated the piece that does not depend on $\pi_d$ as a constant as it does not contribute to the $\delta N$ computation. Note that, although $N_{SR}$ is also a function of $\phi_c$, its dependence is fixed as it is a parameter of the model. In other words, instead of $\phi$ as an independent variable to be varied for the computation of $\delta N$, the independent variable is $\pi_d$ in the present case.

Adding $N_{USR}$ in \eqref{USR-N} and $N_{SR}$ in \eqref{SR-N} together, we obtain the total number of e-folds from the USR phase to the end of inflation specified by $\phi=\phi_f$.
Removing the index $i$ from the initial values $\phi_i$ and $\pi_i$, we obtain
\begin{equation}
    \begin{aligned}
        N(\phi,\pi;\phi_f) &= N_{USR}+N_{SR}
        =\frac{1}{\eta_V} \log\( -2\eta_V \pi_d-6\sqrt{2\epsilon_V} \)
        + \frac{1}{3}\log \left( \frac{\pi}{\pi_c} \right) +\text{const.} ~,
        \label{total N of USR to SR}
    \end{aligned}
\end{equation}
where $\pi_d$ is a function of $\pi_c$ given by \eqref{picpid}, and $\pi_c$ is a function
of the initial $\phi$ and $\pi$ as $\pi_c=\pi+3(\phi-\phi_c)$ through \eqref{pic}.
The constant part refers to the terms without $\phi$ and $\pi$ dependence.

To illustrate the effects of the step, it is helpful to make a comparison with the case without a step. When $\Delta V = 0$, or equivalently $g=\pi_d/\pi_c=1$, we have $\pi_c=\pi_d$ and the above analysis simply goes back to the smooth transition (for $h=0$) or the sharp transition (for $h<0$) discussed in Ref. \cite{Cai:2017bxr}.
The effect of a step is to render $\pi_d$ a nonlinear function of $\pi_c$ given by \eqref{picpid}. As we shall see below, this gives rise to a distinctive non-perturbative feature in the distribution function of the curvature perturbation. 

Finally, let us consider the number of e-folds when the  initial value of $\phi$ is after the step, $\phi<\phi_c$. In this case we have the standard slow-roll result,
\begin{equation}
    \begin{aligned}
      N(\phi;\phi_f) &= \frac{1}{\eta_V} \log \[ 1+\frac{\eta_V}{\sqrt{2\epsilon_V}}(\phi_f-\phi) \] +\text{const.} ~,
      \label{relaxationN}
   \end{aligned}
\end{equation}
where the validity of the expansion in potential \eqref{eq:potential} implies
the condition $\eta_V/\sqrt{2\epsilon_V}(\phi_f-\phi)\ll1$.

\subsubsection{Off-attractor trajectories}
\label{sec:off}

\begin{figure}[htb]
    \centering
    \includegraphics[scale =0.25]{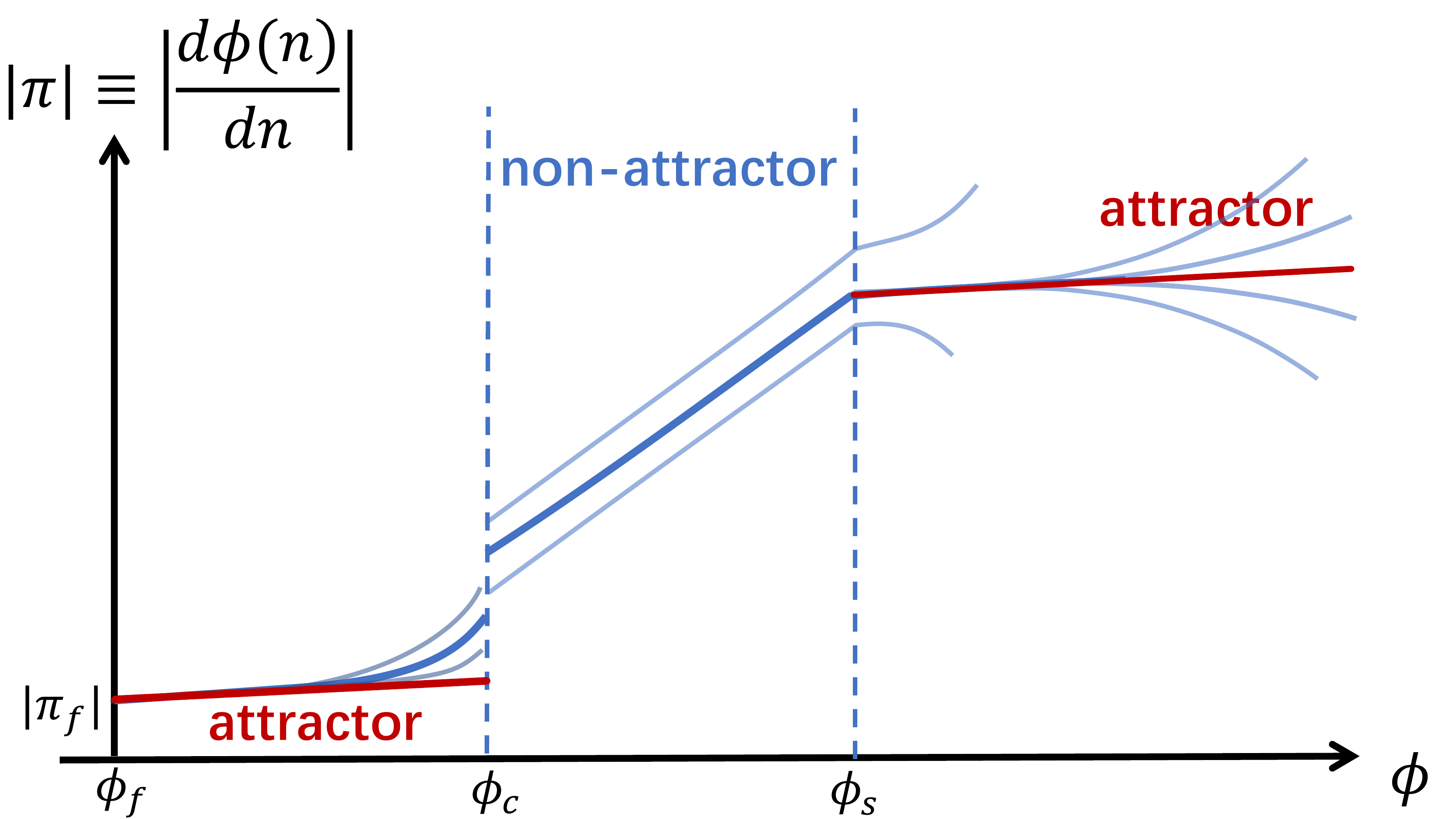}
    \caption{Off-attractor trajectories in the phase diagram $(\phi,\abs{\pi})$ of the SR-USR-SR transition. Slow-roll attractors are depicted in red line. The thick blue line describes the base trajectory while the thin blue lines represent off-attractor trajectories.}
    \label{fig:phase_diagram USR}
\end{figure}
In a more realistic model, an initial SR phase is expected before the USR stage. This corresponds to attaching a SR potential at $\phi>\phi_s$ to the right of the USR potential shown in figure \ref{fig:step} (see also the potential in \ref{fig:potential draft}). The model is similar to the inflection-point inflation, which have been extensively discussed in literature as one of possible mechanisms for producing PBHs. 

Now, let us consider the role of off-attractor trajectories at an initial SR stage prior to the USR stage. At first glance, it seems that the background evolution is fully determined by the value of $\phi$ because the universe must have already arrived at the attractor stage during the initial slow-roll stage. Then the momentum $\pi$ at the beginning of the USR stage, which is the one at the end of the first SR phase, is fixed by the attractor solution $\pi(n_s)=\pi_s$. As a result, although subsequent stages may have non-attractor stages, $\pi$ can still be expressed as a function of the inflaton $\phi$ like in the conventional slow-roll inflation. 

Thus, one might conclude that there exits a unique phase-space trajectory, 
as shown by the thick blue line in figure \ref{fig:phase_diagram USR}, just like the conventional slow-roll inflation, and the non-attractor nature of the trajectory at the intermediate stage would not affect the properties of the perturbation. Let us name this trajectory \textit{the base trajectory}.

A crucial point that has been missed in the above argument is the effect of \textit{off-attractor trajectories} near the end of the first SR stage. 
These trajectories are always present when we consider possible deviations of $\phi$ and $\pi$. At the SR stage, they can be normally neglected, as any deviations from the attractor trajectory vanishes within a couple of e-folds and the SR evolution is quickly recovered.

But the situation may become different when there is a transition. In the SR-USR transition around $\phi_s$, quantum fluctuations may take the trajectory away to an off-attractor one, and some of these off-attractor trajectories may not have enough time to converge to the slow-roll attractor before the USR phase starts. These trajectories, which are shown by light blue curves in figure \ref{fig:phase_diagram USR}, deviate from the base trajectory. Hence, we may have the field velocity $\pi(n_s) \neq \pi_s $ at $\phi_s$, which depends on the off-attractor field velocity $\pi$ at the initial SR stage.
Since $\pi(n_s)$ provides the initial condition for the subsequent stages, this $\pi$-dependence becomes a crucial factor for the whole evolution of the system. In Appendix \ref{App: pi-dependence before the step}, we provide detailed computations and clarify the $\pi$-dependence of the background solution by solving the full dynamics of the initial SR phase.

Apparently, the off-attractor behaviour around the transition plays an important role in the $\delta N$ analysis, as the background evolution of inflation cannot be fully determined by the base trajectory and the initial $\pi$-dependence should also be taken into account. Namely, when computing $\delta N$, it is crucial to include the $\pi$-dependence as well as the $\phi$-dependence.


Interestingly, we find that the above analysis of the off-attractor behaviour may also apply to transitions without an intermediate USR stage. We shall consider such a case, that is, an SR-SR transition with an upward step in Section \ref{sec:SR to SR}.

\subsubsection{Local non-Gaussianity from the $\delta N$ formalism}\label{sec:deltaN}

As has been mentioned previously, the original USR model with a constant potential is able to generate $\mathcal{O}(1)$ local non-Gaussianity of primordial curvature perturbations within the framework of single-field inflation.
There have been various methods to derive this result in the literature, such as the in-in formalism \cite{Namjoo:2012aa, Chen:2013aj, Chen:2013eea, Cai:2017bxr},  the background wave method \cite{Bravo:2017wyw}, the operator product expansion \cite{Finelli:2017fml}.
Among them, the $\delta N$ formalism provides a simple and intuitive way.
This method is based on the separate universe assumption and successfully captures the non-linear evolution of perturbations on super-Hubble scales \cite{Salopek:1990jq, Sasaki:1995aw, Starobinsky:1986fxa, Sasaki:1998ug, Lyth:2004gb, Lee:2005bb, Lyth:2005fi}.
Previously this approach has also been used in the analysis of both smooth and sharp USR-SR transitions \cite{Cai:2017bxr}.
In this section, we apply the $\delta N$ computation to our two models to compute the local non-Gaussianity generated from the transition with an upward step.

First, we consider the USR-SR transition. At the USR stage, the inflaton fluctuation $\delta \phi$ behaves like a massless scalar with no interactions in pure de Sitter space, and its probability distribution is Gaussian.
For the perturbation mode which exits the Hubble radius during USR at $n=n_i$,
we can simply use the number of e-folds derived in \eqref{total N of USR to SR}, and get the following $\delta N$ formula from the definition:
\begin{align}
   \delta N  = N(\phi_i+\delta\phi, \pi_i+\delta\pi) - N(\phi_i, \pi_i)\simeq \frac{1}{\eta_V}\log \left[1 + \frac{2\eta_V\delta\pi_d}{6\sqrt{2\epsilon_V} + 2\eta_V \pi_d}\right]
    - \frac{1}{3}\log \left[ 1 + \frac{3\delta \phi}{\pi_c} \right] ~, \label{exact_calR}
\end{align}
where we have neglected $\delta\pi$ due to its exponential decay on super-Hubble scales, and $\delta \pi_d$ is given by
\begin{align}
    \delta \pi_d
    =\pi_d\left[\sqrt{1+\frac{6}{g}\frac{\delta\phi}{\pi_d} + 9\left(\frac{\delta\phi}{\pi_d}\right)^2} - 1\right] ~.
    \label{delta pi_d as a function of delta phi}
\end{align}

With $\mathcal{R}=\delta N$, the formula in \eqref{exact_calR} yields a non-perturbative mapping between the curvature perturbation $\mathcal{R}$ and the Gaussian field fluctuation $\delta\phi$.
This relation not only captures the case when $|\mathcal{R}|$ is small, but also remains valid for the tail of the probability distribution with rare but large $|\mathcal{R}|$.
We shall elaborate on the implications of the non-Gaussian tail in the next section. Before that, let us first perform the analysis in the perturbative regime $|\mathcal{R}|\ll 1$.

When perturbation theory is valid, we can expand the formula in $\delta \phi$. Up to second order we get
\begin{align} \label{deltaNpert}
\calR \simeq \[ \frac{6}{g^2 (h+2\eta_V)} - 1 \] \frac{\delta\phi}{\pi_c} +  \[ 9\frac{  \left(g^2-1\right) h+2 \eta_V\left(g^2-2\right)}{g^4 (h+2 \eta_V)^2}+\frac{3}{2}\] \(\frac{\delta\phi}{\pi_c}\)^2 .
\end{align}
The linear term corresponds to the Gaussian part
\begin{align}
\calR_{\text{G}} \equiv \[ \frac{6}{g^2 (h+2\eta_V)} - 1 \] \frac{\delta\phi}{\pi_c} \simeq \(\frac{6}{g^2 h} - 1 \) \frac{\delta\phi}{\pi_c}\,.
\label{Rgauss}
\end{align}
This part of the contribution determines the amplitude of the power spectrum, the evaluation of which is deferred to Section \ref{matching}.
Notice that, the second term in the bracket is the size of $\calR$ at the end of the USR phase, while the first term is related to the  step transition.
If $g^2|h|\gg 6$, the upward step does not change the USR result; but for $g^2|h|\ll 6$ the effect of the step becomes dominant. This can happen when the step size is big enough to significantly reduce the kinetic energy of the inflaton, {\it i.e.} $g^2\ll 1$. In that limit we have $\calR_{\text{G}} \simeq {\delta\phi}/({g\sqrt{2\epsilon_V})} $.

\begin{figure}[htb]
    \centering
    \includegraphics[scale =0.8]{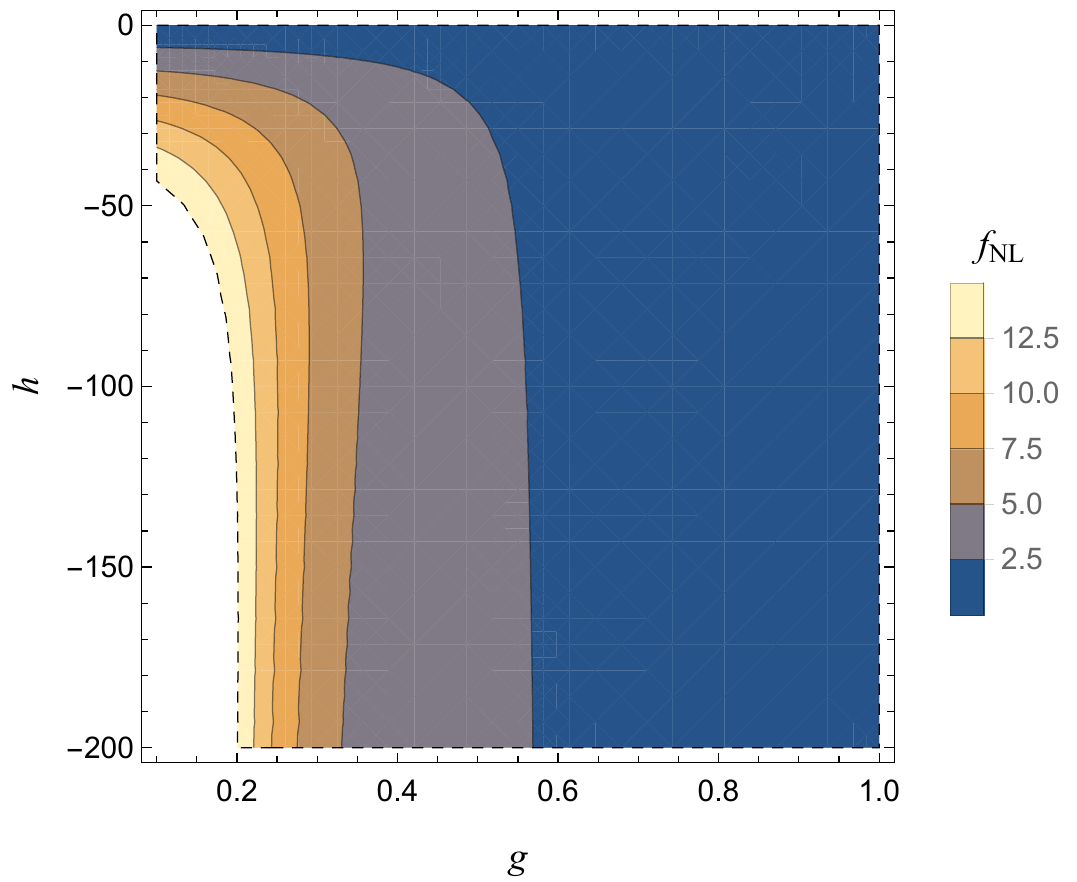}
    \caption{The size of $\fnl$ in the  $(g, h)$ parameter space.(The white region corresponds to $f_{\rm NL}> 15$.)}
    \label{fig:fnl}
\end{figure}

Following the standard treatment, the leading order local non-Gaussianity generated in this model is given by
\begin{align}
    f_{\text{NL}}&= \frac{5}{6} \frac{\partial^2 N}{\partial \phi^2} \left/ \(\frac{\partial N}{\partial \phi}\)^2 \right.
\simeq\frac{5\left(g^4h^2 + 6g^2h-6h \right)}
    {2(6-g^2h)^2}
     \label{fNL of step slow-roll} ~,
\end{align}
where we have ignored the correction of $\mathcal{O}(\eta_V)$ to simplify the result.
Again the final result depends on  two independent parameters $g$ and $h$. In general $h\simeq -6\pi_e/\pi_d$ can be any negative value, while $0 < g \leq 1$ reflects how significant the effect of the upward step is. The size of $\fnl$ is shown by the contour plot of the $(g, h)$ parameter space in figure \ref{fig:fnl}.
In the following, we study two particular cases.

For $g=1$, which means no step, we reproduce the results of the smooth and sharp transitions discussed in Ref. \cite{Cai:2017bxr}.
One finds $\fnl=5h^2/[2(6-h)^2] $, and its maximum value is $\fnl=5/2$ in the limit of $h\rightarrow -\infty$, which corresponds to an infinitely sharp transition. On the other hand,  for $g\ll 1$ where the step becomes important, it is possible to have a large local non-Gaussianity with $\fnl\gg 5/2$, as seen in figure \ref{fig:fnl}.
In the limit $g^2|h|\ll 6$ where the contribution from the step dominates the curvature perturbation, as seen from Eq.~(\ref{Rgauss}), $f_{\rm NL}$ Eq.~\eqref{fNL of step slow-roll} can be approximated as
\begin{equation} \label{eq:fnl}
    f_{\rm NL} \simeq \frac{5}{12}|h| ~.
\end{equation}
Thus, as long as $g^2|h|\ll1$, $\fnl$ is determined by the value of $h$.
This result shows that even in the framework of canonical single-field inflation, we can achieve large local non-Gaussianity with $\fnl \gg \mathcal{O}(1)$, which easily breaks the upper limit of $5/2$ in sharp and smooth transitions. This serves as an intriguing counterexample that violates Maldacena's consistency relation in single-field inflation~\cite{Maldacena:2002vr}.

At first sight, it also seems possible to have infinitely large local $\fnl$ in this model by finely tuning the model parameters. However, we should note that the perturbative treatment breaks down when $|\fnl \calR| \gtrsim 1$. 
In particular this corresponds to the situation where Taylor expansion of the $\delta N$ formula in \eqref{deltaNpert} becomes invalid, and thus we need to reconsider the full expression in \eqref{exact_calR}. 
An intriguing fact is that, as the validity of the perturbative expansion is controlled by the amplitude of $|\fnl \calR|\ll1$, we may encounter the non-perturbative regime when $|\fnl \calR|={\cal O}(1)$, that is, when we look at the rare and large perturbations at the tail of the probability distribution even if $|\fnl|\ll1$. 
A detailed analysis of the non-Gaussian tail is discussed in the next section.

Finally let us consider the perturbation modes which exit the Hubble radius after the step transition. 
With the number of e-folds in \eqref{relaxationN}, we derive
\be
\delta N = \frac{\delta\phi}{\sqrt{2\epsilon_V}} - \frac{\eta_V}{2} \(\frac{\delta\phi}{\sqrt{2\epsilon_V}}\)^2 + ... ~.
\ee
This gives us the standard results of slow-roll inflation for the amplitude of curvature perturbation with $\fnl=-5\eta_V/12$. 
It may be noted that even though there is a non-attractor stage where slow-roll conditions are violated, the curvature perturbation is still given by
the slow-roll attractor formula as if there were no non-attractor stage~\cite{Leach:2001zf}.
As a result, the local non-Gaussianity remains the same as that for the slow-roll case.\footnote{For these small-scale modes, there may be large intrinsic non-Gaussianities in the inflaton fluctuations (hence they would not be of the local type) due to the discontinuity in the potential. This part of the analysis is beyond the scope of the present paper.}

\subsection{SR-SR transition with an upward step}\label{sec:SR to SR}

So far, we have focused on the upward-step transition from an USR stage.
In this subsection we consider an upward step in the SR-SR transition. 
The model is the one proposed in \cite{Cai:2021zsp}. Here we analyse it in more detail.

\begin{figure}[htb]
    \centering
    \includegraphics[scale = 0.4]{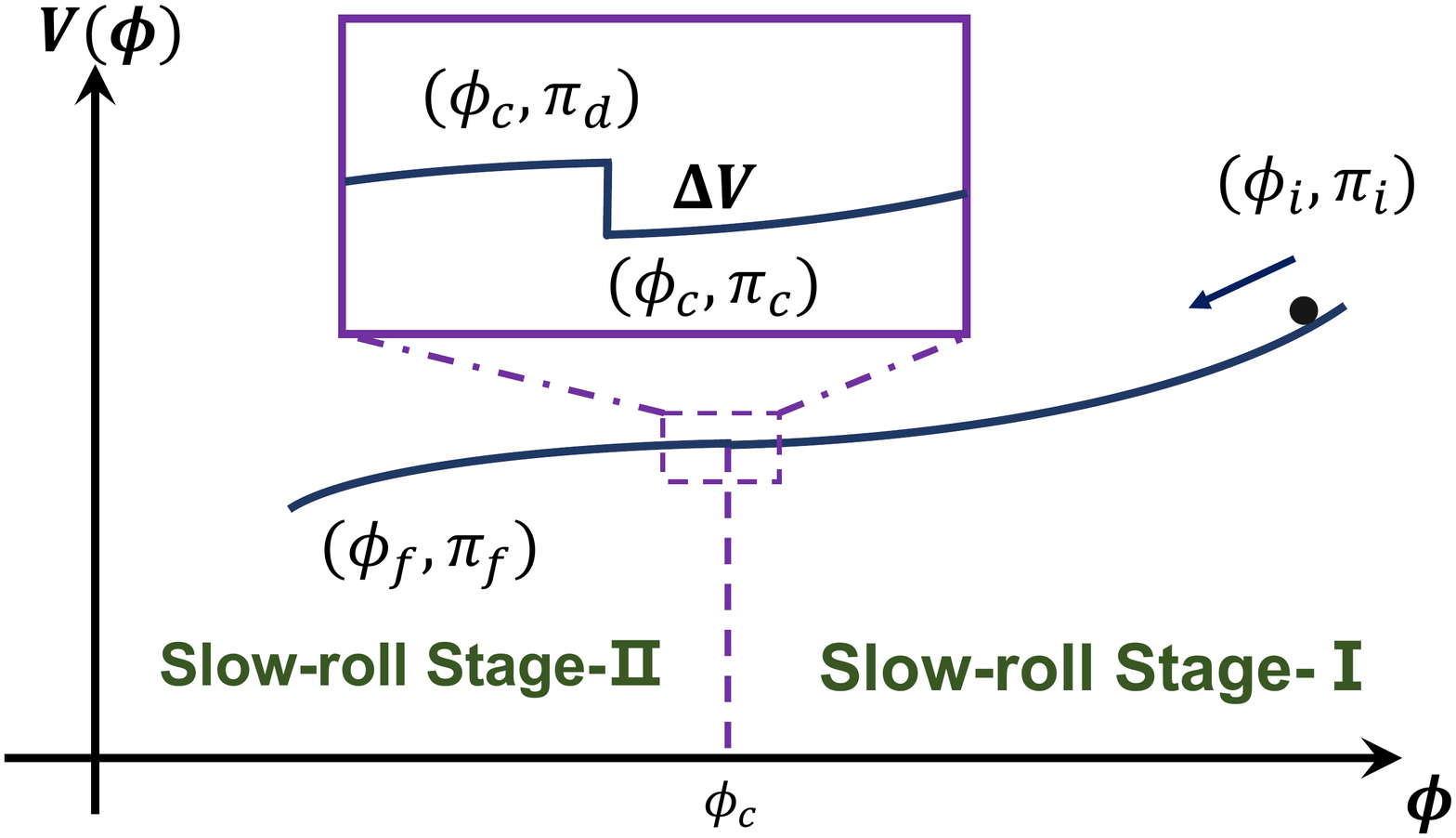}
    \caption{A sketch of the inflaton potential for the SR-SR transition, where two stages of slow-roll inflation are connected by an upward step.}
    \label{fig:SR to SR}
\end{figure}

We consider the two-stage slow-roll inflation with two distinct slow-roll potentials joined at $\phi=\phi_c$ with an upward step. The inflaton initially rolls on a slow-roll potential from $\phi> \phi_c$ (Stage-I), and climbs up an upward step at $\phi_c$, as shown in figure \ref{fig:SR to SR}, and rolls down on the second slow-roll potential after a short relaxation time (Stage-II). 
To solve the background dynamics of this model, we first parametrize the two slow-roll potentials as
\begin{align}
    V(\phi) = 
        \begin{dcases}
            V_0 \left[ 1+ \sqrt{2\epsilon_{\rom{1}}}\left(\phi -\phi_c \right) + \frac{1}{2}\eta_{\rom{1}}\left(\phi -\phi_c \right)^2\right] ~,\quad \phi \geq \phi_c\\
            \left(V_0 + \Delta V \right)\left[ 1 + \sqrt{2\epsilon_{\rom{2}}}\left(\phi -\phi_c \right) + \frac{1}{2}\eta_{\rom{2}}\left(\phi -\phi_c \right)^2\right] ~,\quad \phi < \phi_c
        \end{dcases} \qquad ,
\end{align}
where $\Delta V $ is the size of the step, and $\epsilon_{I,II}$ and $\eta_{I,II}$ are the Stage-I and Stage-II slow-roll parameters defined at $\phi_c$.

Following Sec. \ref{sec:usr-sr}, we denote the field velocities before and after the step by $\pi_c$ and $\pi_d$, respectively.
For the  evolution in Stage-II, the background solutions are the same with what found in Sec. \ref{sec:usr-sr}. We may simply use the results there, by replacing $\epsilon_V\rightarrow\epsilon_{II}$ and $\eta_V\rightarrow\eta_{II}$.
It is a bit nontrivial to derive the full evolution in Stage-I with arbitrary initial conditions. 
This background analysis is essential for the studies of long wavelength perturbations which exit the horizon during Stage-I.
We leave the detailed computation in Appendix \ref{App: pi-dependence before the step}.
Here we present a simple analysis with major results.
For the Stage-I evolution with non-slow-roll initial conditions,
there can be two different situations:

\begin{itemize}
\item[(a)] For the initial condition $(\phi_i,\pi_i)$ sufficiently far from the step, the trajectory quickly approaches the slow-roll trajectory, and starts following the attractor evolution. Thus the conventional
slow-roll approximations can still apply, and we find
\begin{align}\label{slow-roll eq in Stage-I}
    \dv{\phi}{n} + \sqrt{2 \epsilon_{\rom{1}}} + \eta_{\rom{1}}\left(\phi-\phi_{c}\right) = 0 ~,\quad \phi_c < \phi~.
\end{align}
At the end of Stage-I, the field momentum is given by $\pi_c = -\sqrt{2 \epsilon_{\rom{1}}}$, which is fully fixed by the shape of the slow-roll potential at Stage-I. This corresponds to the base trajectory whose dynamics is determined regardless of initial conditions.
Then the number of e-folds in this stage $N_{\rom{1}}$ can be directly solved  from \eqref{slow-roll eq in Stage-I}:
\begin{align}
    N_{\rom{1}} = n_c - n_i = \frac{1}{\eta_{\rom{1}}}\log\left[ 1 + \frac{\eta_{\rom{1}}}{\sqrt{2\epsilon_{\rom{1}}}}\left(\phi_i - \phi_c\right) \right] ~.\label{N of the first stage }
\end{align}
After the step, $N_{\rom{2}} \equiv n_f - n_c $ is essentially the same as \eqref{SR-N}. So in this case the total  number  of e-folds is given by
\begin{equation}\label{total N for a}
    \begin{aligned}
        N_{\text{total}} &= N_{\rom{1}} + N_{\rom{2}}\\
    &\simeq
    \frac{1}{\sqrt{2\epsilon_{\rom{1}}}}\left( \phi_i - \phi_c \right)
    +
    \frac{1}{\eta_{\rom{2}}}\log\left[ 
    -2\eta_{\rom{2}}\pi_d - 6\sqrt{2\epsilon_{\rom{2}}} \right] + \text{constant} ~.
    \end{aligned}
\end{equation}
Here we notice that there is a major difference with the USR-SR transition discussed in Sec. \ref{sec:usr-sr}.
For the total number of e-folds $N$ in \eqref{total N of USR to SR}, $\pi_d$ is a function of the initial conditions $(\phi_i,\pi_i)$ in the USR stage. However, here the field momentum $\pi_d$ 
is fully fixed by $\pi_c = -\sqrt{2\epsilon_{\rom{1}}}$ and the size of the step. Therefore in this case, $N_{\rom{2}}$ can also be seen as a constant, which is independent of initial conditions.
For perturbation modes which exit the  horizon at $\phi = \phi_i$,
the $\delta N$ formula simply gives us the standard slow-roll result,
\begin{equation}\label{slow-roll result}
    \begin{aligned}
        \calR =\delta N=
        -\frac{\delta \phi}{\sqrt{2\epsilon_{\rom{1}}}} ~.
    \end{aligned}
\end{equation}

\item[(b)] For the initial condition $(\phi_i,\pi_i)$ close to the step,
 there may not be enough time for the trajectory to converge to the slow-roll trajectory before it encounters the step. 
 This corresponds to the off-attractor trajectories we discussed in Sec. \ref{sec:off}. Thus the dependence on $\pi_i$ in the initial condition becomes crucially important.
 
As shown in Appendix \ref{App: pi-dependence before the step}, the off-attractor trajectories demonstrate the non-attractor behaviour just like the USR case,
\begin{align} \label{non-attractor dynamics}
    3\phi(n) + \pi(n) = 3\phi_i + \pi_i~.
\end{align}
Then the number of e-folds $N_I$ until the step is given by
\begin{align}
    N_I =\frac{1}{3} \log\[\frac{\pi_i}{\pi_c}\]= \frac{1}{3} \log\[ \frac{\pi_i}{\pi_i+3(\phi_i-\phi_c)}\].
\end{align}
Adding the number of e-folds of Stage-II to the above, we find
\begin{equation}\label{total N}
    \begin{aligned}
        N_{\text{total}} &= N_{\rom{1}} + N_{\rom{2}}\\
    &\simeq
    \frac{1}{3} \log\[ \frac{\pi_i}{\pi_i+3(\phi_i-\phi_c)}\]
    +
    \frac{1}{\eta_{\rom{2}}}\log\left[ 
    -2\eta_{\rom{2}}\pi_d - 6\sqrt{2\epsilon_{\rom{2}}} \right] + \text{constant} ~.
    \end{aligned}
\end{equation}
Now, contrary to the case (a), $\pi_d$ in this case depends on the initial condition through  \eqref{picpid} and $\pi_c =3(\phi_i-\phi_c) + \pi_i $. For the perturbation modes which exit horizon close to the end of Stage-I, we find the $\delta N$ result identical to the one for the USR-SR transition,
\begin{equation} \label{near step modes}
    \begin{aligned}
        \calR &= N(\phi_i+\delta\phi, \pi_i+\delta\pi) - N(\phi_i, \pi_i)\\
        &\simeq
        - \frac{1}{3}\log \left[ 1 + \frac{3\delta \phi}{\pi_c} \right] + 
        \frac{1}{\eta_{\rom{2}}}\log \qty[1 + \frac{2\eta_{\rom{2}}\delta\pi_d}{6\sqrt{2\epsilon_{\rom{2}}} + 2\eta_{\rom{2}} \pi_d} ]\,,
    \end{aligned}
\end{equation}
where $\delta \pi_d$ is given by \eqref{delta pi_d as a function of delta phi}. 
\end{itemize}

With the above analysis, we now have a good description of the non-Gaussianity generated in the SR-SR transition with an upward step. 
For a broad range of perturbation modes which exit the horizon during Stage-I, the standard slow-roll result in \eqref{slow-roll result} still applies, and thus the local $\fnl$ is slow-roll suppressed. 
However, for a narrow range of modes which exit the horizon right before the step, the off-attractor trajectories become important, and thus we should resort to \eqref{near step modes} to study the nonlinear perturbations. Here we focus on the perturbative regime  $\abs{\calR} \ll 1$,  and leave the discussion of non-Gaussian tails around  $\abs{\calR} \sim 1$ in the next section.
 
The $\delta N$ expansion for the case (b) is the same as that for the USR-SR transition, Eq.~(\ref{deltaNpert}),
\begin{align} 
    \calR \simeq \[ \frac{6}{g^2 (h+2\eta_{\rom{2}})} - 1 \] \frac{\delta\phi}{\pi_c} +  \[ 9\frac{  \left(g^2-1\right) h+2 \eta_{\rom{2}}\left(g^2-2\right)}{g^4 (h+2 \eta_{\rom{2}})^2}+\frac{3}{2}\] \(\frac{\delta\phi}{\pi_c}\)^2\,,
\end{align}
with the linear term being the Gaussian part,
\begin{align}
    \calR_{\text{G}} \equiv \[ \frac{6}{g^2 (h+2\eta_{\rom{2}})} - 1 \] \frac{\delta\phi}{\pi_c} \simeq \(\frac{6}{g^2 h} - 1 \) \frac{\delta\phi}{\pi_c}\,.
\end{align}
Again, as in the case of the USR-SR transition, this part of the contribution determines the amplitude of the power spectrum, which will be evaluated in Sec.~\ref{matching}.
The local non-Gaussianity generated in this model is given by
\begin{align}
        f_{\text{NL}}&= \frac{5}{6} \frac{\partial^2 N}{\partial \phi^2} \left/ \(\frac{\partial N}{\partial \phi}\)^2 \right.
    \simeq\frac{5\left(g^4h^2 + 6g^2h-6h \right)}
        {2(6-g^2h)^2}\,,
\end{align}
which exactly agrees with the one in the USR-SR case given in Eq.~(\ref{fNL of step slow-roll}). Again, in the limit when the effect of the step is significant, $g^2 \ll 1$, the above reduces to $f_{\text{NL}}^{\text{local}} \simeq 5\abs{h}/12$, as given by Eq.~\eqref{eq:fnl} for the USR-SR transition. 

\section{A tale of the non-Gaussian tail}\label{NG tail}

Now we study the probability distribution of the curvature perturbation beyond the perturbative regime. We focus on the behaviour of the tail due to the upward step transition.

Here we are interested in the local non-Gaussianity, which in general arises when the curvature perturbation is a function of a Gaussian random field $\calR_{\text{G}}$ as
\be \label{fullcalR}
\calR({\bf x}) = f\(\calR_{\text{G}}({\bf x}) \) .
\ee
This function is arbitrary in principle apart from the condition that $f(0)=0$. Further, if we assume that $\calR$ is linear in $\calR_{\text{G}}$ in the limit $\calR_{\text{G}}\to0$, we can also rescale $\calR_{\text{G}}$ such that $f'(0)=1$.
Then, in perturbation theory, we have the Taylor expansion as $|\calR_{\text{G}}|\ll 1$,
\be \label{taylorR}
\calR({\bf x}) = \calR_{\text{G}}({\bf x}) + \frac{1}{2}f''\(0\) \calR_{\text{G}}({\bf x})^2 + ...
\ee
which gives us $\fnl= 5f''(0)/6$.
Note that the nonlinear relation \eqref{fullcalR} and its perturbative expansion \eqref{taylorR} are exactly what we obtain from the $\delta N$ formalism.
When we look at the probability distribution of the curvature perturbation, normally a positive $\fnl$ means that the probability is higher than the Gaussian distribution for positive values of $\calR$. 
However this naive expectation is only valid in the perturbative regime. As we have argued previously, when we consider large values of $|\calR|$, the perturbative expansion \eqref{taylorR} may break down, and we may need a full nonlinear expression of $f\(\calR_{\text{G}}({\bf x}) \)$ to capture non-perturbative effects.

One of well-studied examples in the literature is based on the original USR scenario. From the analytical expression of $N$ in \eqref{USR-N}, one easily finds that the nonlinear mapping  from $\delta\phi$ to $\calR$ is given by  $\calR = \delta N= -\frac{1}{3}\log \( 1+3 \delta\phi /\pi_c \right) $. In the perturbative regime, this yields $\fnl=5/2$, while the non-Gaussian probability distribution is enhanced at large $\calR$.
Thus in this example, one finds that a positive $\fnl$ leads to more distribution of large perturbations on the tail, which agrees with the naive expectation from the perturbative analysis.

Now let us go back to the USR-SR transition with an upward step. The non-perturbative relation between $\calR$ and $\delta \phi$ is given in \eqref{exact_calR}. We may further simplify the expression by
taking the $\eta_V\rightarrow 0$ limit,
\begin{equation}
    \calR \simeq
     -\frac{1}{3}\log\left( 1 + 3 \frac{\delta \phi}{\pi_c} \right)+\frac{2}{h}\left[ \sqrt{1+\frac{6}{g}\frac{\delta\phi}{\pi_d} + 9\left(\frac{\delta\phi}{\pi_d}\right)^2} - 1\right] ~. \label{calR_phi}
\end{equation}
Here the first term with a logarithmic function is contribution from the USR stage, whose effects on the probability distribution have been discussed in the literature \cite{Franciolini:2018vbk, Biagetti:2018pjj, Atal:2018neu, Passaglia:2018ixg, Atal:2019cdz, Atal:2019erb, Taoso:2021uvl, Biagetti:2021eep, Davies:2021loj}.
The second term with the square root is the new contribution caused by the upward step.
Here let us focus on the case when $g^2|h|\ll 1$ where the curvature perturbation is dominated by the effect of the step.
In this parameter regime, one can neglect the logarithmic term to obtain
\begin{align}
    \calR &\simeq -\frac{2}{\abs{h}}\left[\sqrt{1-|h|\calR_{\text{G}} } - 1\right] ~,\label{approximate solutioin of calR}
\end{align}
where $\calR_{\text{G}}=6\,\delta\phi/(gh\pi_d) $ is the Gaussian part of the curvature perturbation when $|\calR| \ll 1$.
We immediately notice the presence of a cutoff at $\calR=2/|h|$. This is a genuine non-perturbative effect. Physically, it corresponds to the situation when the inflaton fluctuation is too large to render the field momentum too small to climb the upward step. The formula~\eqref{approximate solutioin of calR} is one of our major results of this paper.

It should be noted that the sharp cutoff in $\calR$ is obtained without taking the quantum diffusion effect into account. In particular, we neglected the first term in \eqref{calR_phi}, retaining which would recover the leading order quantum diffusion effect obtained in the stocastic $\delta N$ 
formalism~\cite{Jackson:2022unc,Animali:2022otk,Pattison:2021oen,Firouzjahi:2018vet}.
Thus quantum diffusion may smooth the cutoff and lead to a non-vanishing tail of probability distribution function (PDF) at $\calR > 2/|h|$ as it allows the inflaton to
climb the step.
It may become important if the step height is low, $g^2=O(1)$, and the number of $e$-folds of the USR stage is large, $N_{USR}\gg1$.
However, in our case of interest, we have $g^2\ll1$ and $N_{USR}=O(1)$.
Hence its effect is exponentially suppressed, i.e. the step height is so high that the probability of the inflaton to climb up the step by quantum diffusion is 
exponentially small. This justifies the approximate solution given by \eqref{approximate solutioin of calR}.

One might worry what would happen to a region which fails to climbe up the step due to quantum fluctuations. In this case, the region trapped at the bottom
will keep inflating forever, and the probability of the whole region to climb up the step becomes completely negligible. However, seen from outside of the region, 
it will eventually collapse when the potential energy of the surrounding region becomes smaller than that of the trapped region. This means the formation of a black hole.
We will discuss it separately in Section \ref{Subsec: trapping PBHs}.

\begin{figure}[H]
  \centering
  \includegraphics[scale = 0.7]{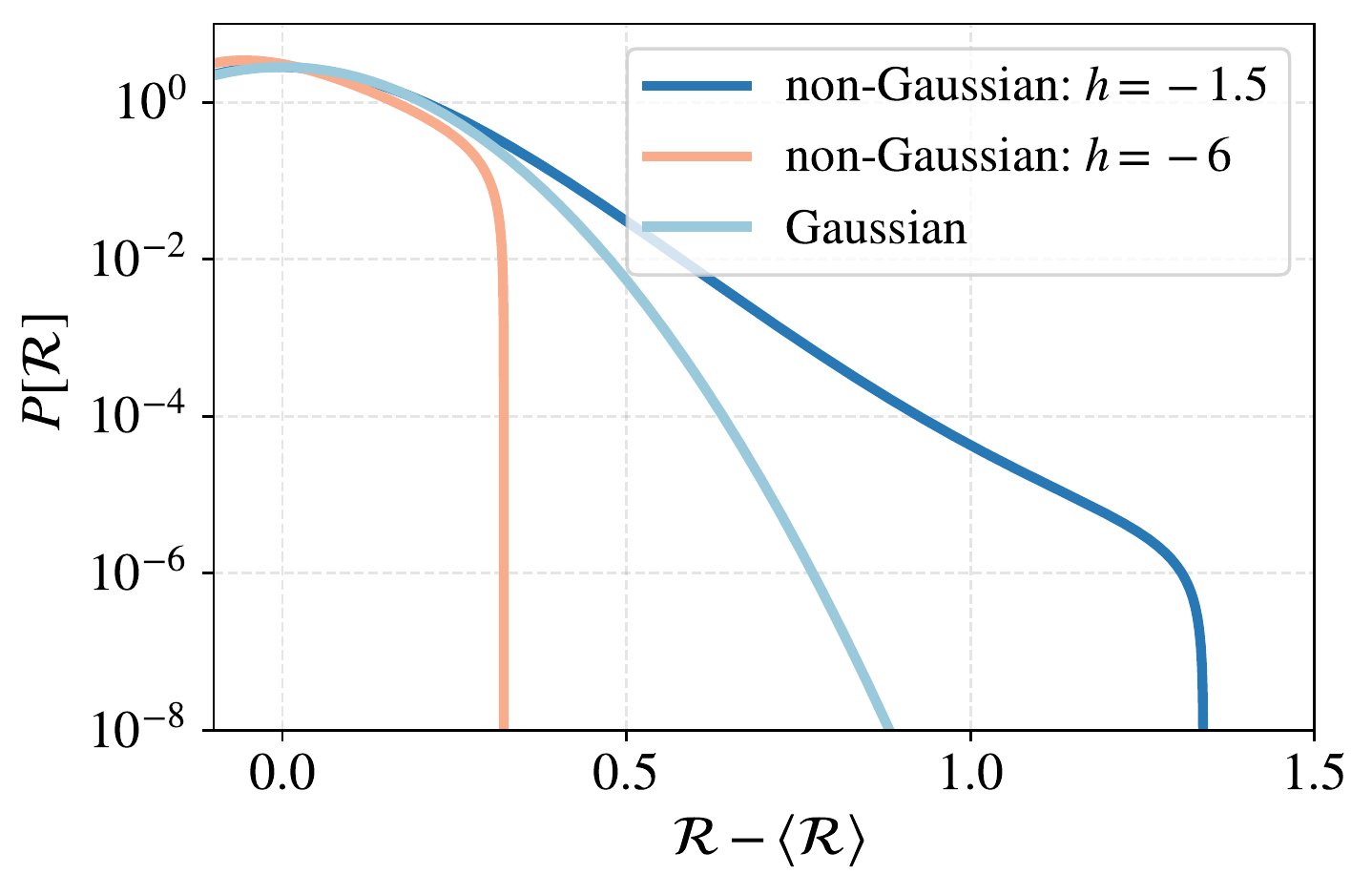}
 \caption{The PDF in \eqref{pdfnon-G} compared with a Gaussian distribution.}
    \label{fig:pdf}
\end{figure}

To see the non-Gaussian feature of the tail more clearly, let us compute the probability distribution function (PDF) of $\cal R$.  
For the Gaussian perturbation $\calR_{\text{G}}$, the PDF is given by
\be \label{Gaussian}
 P[\calR_{\text{G}}] = \frac{1}{\sqrt{2\pi} \sigma_{\calR}} e^{-{\calR_{\text{G}}^2}\left/{2\sigma_{\calR}^2}\right.} ,
\ee
where $\sigma_{\calR}^2$ is the variance of the Gaussian perturbation $\calR_{\text{G}}$,
\be
 \sigma_{\calR}^2 = \int d\log k \mathcal{P}_{\calR_{\text{G}}}(k)\,,
\ee
where we have implicitly assumed that the power spectrum of $\calR_\text{G}$ is peaked at a certain scale, say at $k=k_*$, and ignored the contribution from the spectrum far away from the peak. 
From Eq.~\eqref{approximate solutioin of calR}, the PDF of the curvature perturbation is derived as
\be \label{pdfnon-G}
 P[\calR]=P[\calR_{\text{G}}]  \left| \frac{d \calR_{\text{G}}}{d \calR} \right|
= \frac{2-\abs{h}\calR}{\Omega}\exp\left[ -\frac{\calR^2(4-\abs{h}\calR)^2}{32\sigma^{2}_{\calR}} \right]\,;\qquad\calR\leq\frac{2}{\abs{h}}\,~,
\ee
where the $\Omega$ is a normalization coefficient :
\begin{align}
    \Omega \equiv \sqrt{2\pi\sigma^{2}_{\calR}}\left[1+ \rm{Erf}\left(\frac{1}{\abs{h}\sqrt{2\sigma^2_{\calR}}}\right) \right]~.
\end{align}
The comparison between this PDF and the Gaussian one is shown in figure \ref{fig:pdf}, where $\sigma_\calR^2=0.02$ for a demonstration. It may be noted that the expectation value $\langle\cal R\rangle$ is non-zero. Nevertheless it is found to be too small to be seen in the figure for the current choice of $\sigma_\calR$. As we can see, in the perturbative regime the probability distribution behaves like Gaussian. But as $|\calR|$ becomes large, the deviation from the Gaussian distribution becomes more and more significant, and there appears a sudden cutoff at $\calR = 2/|h|$. 

As we see from Eq.~(\ref{approximate solutioin of calR}),
the parameter $h$ determines the non-Gaussian nature of the PDF in the limit $g^2|h|\ll1$.
Recalling that $h$ is inversely proportional to the ratio of the field velocity after the step and that for the slow-roll attractor $h=-6\pi_e/ \pi_d$, it can take any negative number. 
Depending on the values of $h$, there are at least two interesting cases:

\begin{itemize}
\item For $|h|\gg1$, Eq.~(\ref{approximate solutioin of calR}) implies that there are large deviations from the Gaussian PDF even at $|\calR|\ll1$, and the upper limit of $\calR$ becomes small.
Namely, the probability of large $\calR$ vanishes completely. 
Thus we find a highly suppressed non-Gaussian tail at $\calR\lesssim 2/|h|$, as shown by the orange curve in figure \ref{fig:pdf}.

\item For $|h| \lesssim \order{1}$, the deviation from the Gaussian distribution at $|\calR|\ll1$ is small. In terms of $\fnl$, we have $\fnl=5|h|/12\lesssim1$. On the other hand, the tail of the distribution at large $\calR$ is significantly enhanced, up to its upper limit $\calR=2/|h|\gtrsim1$, where there is a sharp cutoff. An example of this case is shown by the dark blue curve in figure \ref{fig:pdf}. This rather counter-intuitive result provides a clear demonstration that the sizes of the perturbative non-Gaussianity and the non-Gaussianity at the tail are not necessarily related to each other.
\end{itemize}


\section{Power spectrum and implications to the PBH formation}\label{Inflection point}

In this section, we first compute the curvature perturbation power spectrum, and 
study how the non-Gaussian tail may affect the formation of PBHs. 
%
%
A concrete, more realistic inflation model with an upward step transition can be constructed within the context of inflation with an inflection-point potential, 
which has been extensively studied in the literature \cite{Garcia-Bellido:2017mdw, Bhaumik:2019tvl, Anguelova:2020nzl, Karam:2022nym, Germani:2017bcs}.
In this scenario, the inflaton field undergoes SR-USR-SR transitions, and certain modes of the curvature perturbation are enhanced at the USR phase. This leads to an amplified power spectrum on small scales which can efficiently generate PBHs. 

\begin{figure}[htb]
    \centering
    \includegraphics[scale = 0.37]{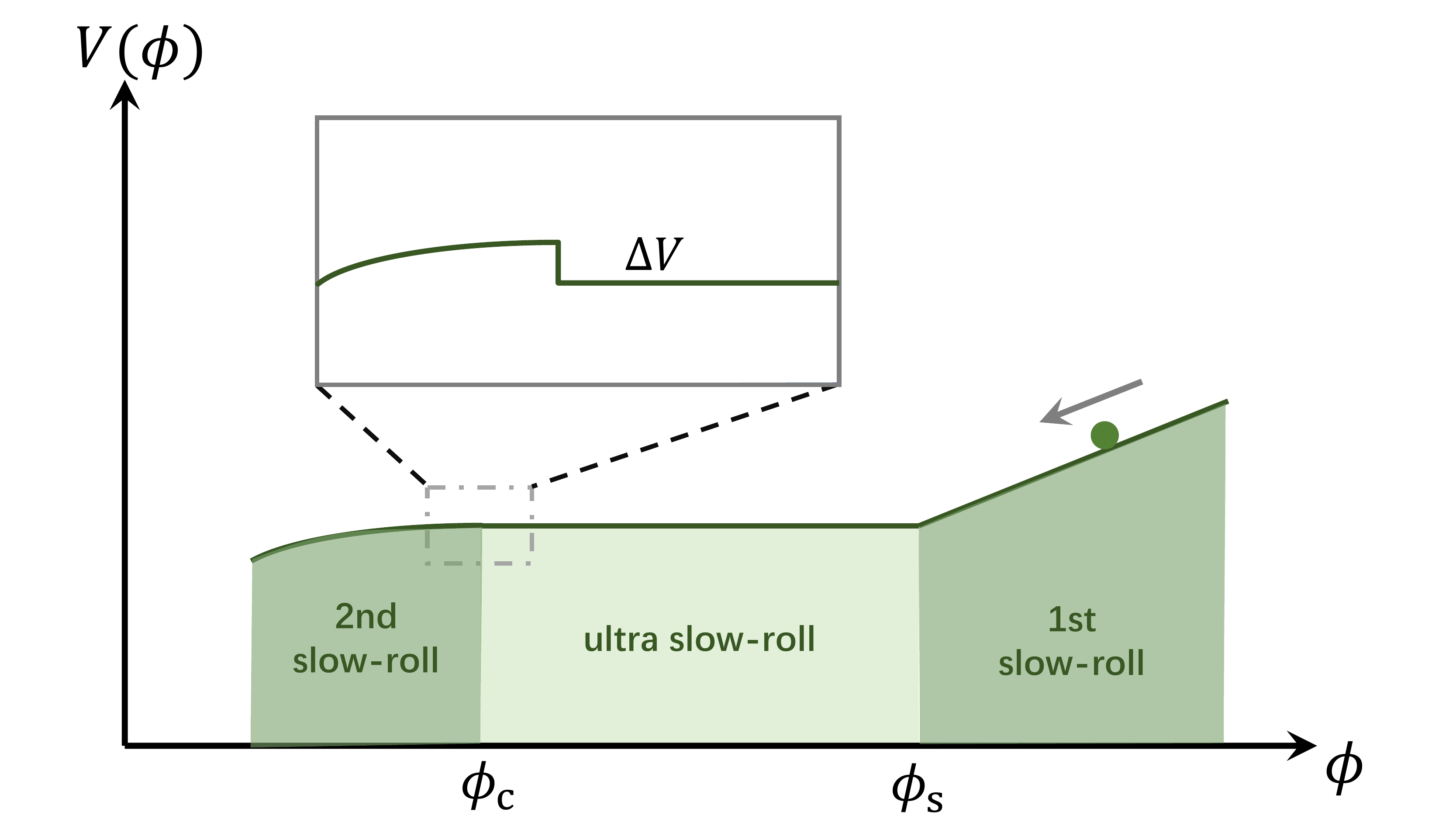}
    \caption{A sketch plot of the inflection-point potential with an upward step. The inflaton starts rolling at a linear potential from $\phi > \phi_s$, and then goes through a flat platform which gives the USR phase and re-enters the slow-roll phase after climbing over the upward step.
    }
    \label{fig:potential draft}
\end{figure}

As an analytic approximation for an inflection-point potential with an upward step, we consider the form,
\begin{align}
    \begin{split}
        V(\phi) = &V_0 \left[ 1 + \sqrt{2\epsilon_S}\left(\phi - \phi_s\right) 
        \Theta\left( \phi-\phi_s\right) \right] \Theta\left( \phi-\phi_c \right) \\
        &+ (V_0+\Delta V)\left[ 1 + \sqrt{2\epsilon_V}\left(\phi - \phi_c\right)
        +\frac{1}{2}\eta_V \left(\phi - \phi_c\right)^2 \right]
        \Theta\left( \phi_c-\phi \right)  ~.
    \end{split}\label{parameterized potential}
\end{align}
Here $\phi_s$ is the field value at the beginning of the USR stage, $\phi_c$ is that at the step, which also marks the end of the USR stage, $\epsilon_S$ is the potential slow-roll parameter of the first SR stage, while $(\epsilon_V,\eta_V)$ are the potential slow-roll parameters of the last SR stage. A sketch of this potential is given in figure~\ref{fig:potential draft}.
Following the notation used in the previous sections, we denote the comoving wavenumber which crosses the Hubble radius at $\phi=\phi_s$ by $k_s$ and that at $\phi_c$ by $k_c$. 

\subsection{The power spectrum from matching}\label{matching}

We study the inflaton field fluctuations and derive the curvature perturbation power spectrum for the potential provided in Eq.~\eqref{parameterized potential}.
We also consider a special limiting case where there is no USR state, that is, $\phi_c = \phi_s$. We present an analytical computation of the power spectrum on large scales $k \lesssim k_s$ first and then of that on small scales $k \sim k_c$.

\subsubsection*{SR-USR-SR Model}

We begin with computing the power spectrum of the extremely long wavelength modes $k \ll k_s$. In this limit, the standard slow-roll result applies, and the Gaussian approximation to the curvature perturbation is valid, with the Gaussian part given by
\begin{equation}
   \mathcal{R}_{\text{G}}= \frac{\delta\phi}{\sqrt{2\epsilon_S}} ~, \quad k \ll k_s ~,
\end{equation}
and the power spectrum by
\begin{equation}
    \mathcal{P}_{\mathcal{R}_{\text{G}}}(k \ll k_s) = \frac{1}{2\epsilon_S}\(\frac{H}{2\pi}\)^2 ~.
\end{equation}
This large scale part may be made to fit the current CMB data with an appropriate choice of $\epsilon_S$. 

For the long wavelength modes $k\lesssim k_s$, the power spectrum shows power-law behaviour with the spectrum index given by
\begin{equation}\label{Spectrum index}
    n_s -1 = \frac{d \log \mathcal{P}_{\mathcal{R}}(k)}{d\log k} \simeq 4\,.
\end{equation}
The details are discussed in Appendix \ref{Inflection point and scaling}.
This power-law behaviour agrees with the typical growth rate of a power spectrum in single-field inflation when there is a stage where the friction-dominated evolution is violated due to a sudden change in the derivative of the potential~\cite{Byrnes:2018txb}.

Next, we consider the spectrum around the step at $\phi_c$. During the USR stage, the mode function for the field fluctuation $\delta\phi$ is given by the standard adiabatic vacuum solution,
\begin{equation}\label{eq:BD-vacuum mode function}
    \delta\phi_k(\tau) = \frac{H}{\sqrt{2 k^3}}(1+ik\tau)e^{-ik\tau}\,; \quad \tau < \tau_c\,,
\end{equation}
where $\tau_c$ is the conformal time at the time of the upward step transition.  

For modes $k \lesssim k_c$, assuming that the perturbative non-Gaussianity is not extremely large (i.e., $|h|$ is not extremely large), we can use the linear $\delta N$ formula to compute the power spectrum. From \eqref{total N of USR to SR}, the Gaussian part of the curvature perturbation is
\begin{equation}
    \calR_{\text{G}} = \frac{\partial N}{\partial\phi} \delta\phi \simeq \(\frac{1}{g}-\frac{gh}{6}\)\frac{\delta\phi}{\sqrt{2\epsilon_V}} ~; \quad k \lesssim k_c ~.
\end{equation}
This yields the power spectrum,
\begin{equation}\label{Long mode power spectrum 1}
\mathcal{P}_{\mathcal{R}_{\text{G}}}(k \lesssim k_c) = \frac{1}{2\epsilon_V}\(\frac{1}{g}-\frac{gh}{6}\)^2
\mathcal{P}_{\delta\phi}(k \lesssim k_c) = \frac{1}{2\epsilon_V g^2}\(1-\frac{g^2h}{6}\)^2 \(\frac{H}{2\pi}\)^2\,.
\end{equation}
Here we mention that the above scale invariant spectrum is obtained under the assumption that the initial stage is USR. If the initial stage is SR, followed by a quickly flattening potential as in the inflection-point inflation, the perturbation modes which exit the Hubble horizon when the inflaton is decelerated by the flattening potential becomes scale-dependent due to the mixing of positive and negative frequencies. In this case, and the spectral index is given by $n_s-1=4$ \cite{Byrnes:2018txb}. The detailed discussion is deferred to Appendix \ref{Inflection point and scaling}.

For modes $k \gtrsim k_c$, Eq.~\eqref{relaxationN} yields
\begin{equation}
    \delta N_{\text{G}} = \frac{\delta\phi}{\sqrt{2\epsilon_V}} ~, \quad k \gtrsim k_c ~.
\end{equation}
Here, however, the spectrum of $\delta\phi$ would not be given by $(H/(2\pi))^2$. Similar to the effect of the SR-USR transition, the upward transition induces the mixing of positive and negative frequencies due to a large change in the derivative of the potential.
The resultant power spectrum takes the form,
\begin{equation}
    \mathcal{P}_{\mathcal{R}_{\text{G}}}(k \gtrsim k_c) = \frac{1}{2\epsilon_V} \mathcal{P}_{\delta\phi}(k \gtrsim k_c) = \frac{1}{2\epsilon_V} \big| \alpha_{k}+\beta_{k} \big|^2 \(\frac{H}{2\pi}\)^2 ~,
\end{equation}
where $\alpha_{k}$ and $\beta_{k}$ are the Bogoliubov coefficients due to the upward step transition. A detailed derivation of the coefficients is given in Appendix \ref{deltaphi gauge}. Here we provide an approximate expression of the power spectrum at the short wavelengths,
\begin{equation}\label{Short wavelength power spectrum}
    \mathcal{P}_{\mathcal{R}_{\text{G}}}(k \gtrsim k_c) \simeq \frac{1}{2 \epsilon_V}\frac{g^4+1+\left(1-g^4\right) \cos (2 k \tau_c)}{2 g^2} \(\frac{H}{2\pi}\)^2 \leq \frac{1}{2 \epsilon_V}\frac{1}{g^2} \(\frac{H}{2\pi}\)^2~.
\end{equation}
Thus the power spectrum scales with the parameter $g$ as
\begin{equation}\label{Power spectrum amplitude scaling}
    \mathcal{P}_{\mathcal{R}}(k) \propto g^{-2}\,.
\end{equation}
Recall that $g$ defined in \eqref{g} parametrizes the size of the upward step; $g$ becomes smaller as the step becomes higher. Thus a higher step amplifies the power spectrum and boosts the production of PBHs.

Another feature in the above spectrum is an oscillating feature with period $2\tau_c$. Notice that the amplitude of these oscillations is almost constant. This result is agreement with the comoving slicing computation where $\mathcal{R}$ is quantized in Appendix \ref{R gauge}. The constancy of the amplitue is an artifact due to our sudden step approximation. Since the step is infinitely sharp, it affects all the comoving wavenumbers up to infinitely large $k$. In reality, the spectrum settles down to the standard slow-roll expression as we shall shortly see below.

\begin{figure}[!htb]
    \centering
    \includegraphics[width=0.45\textwidth]{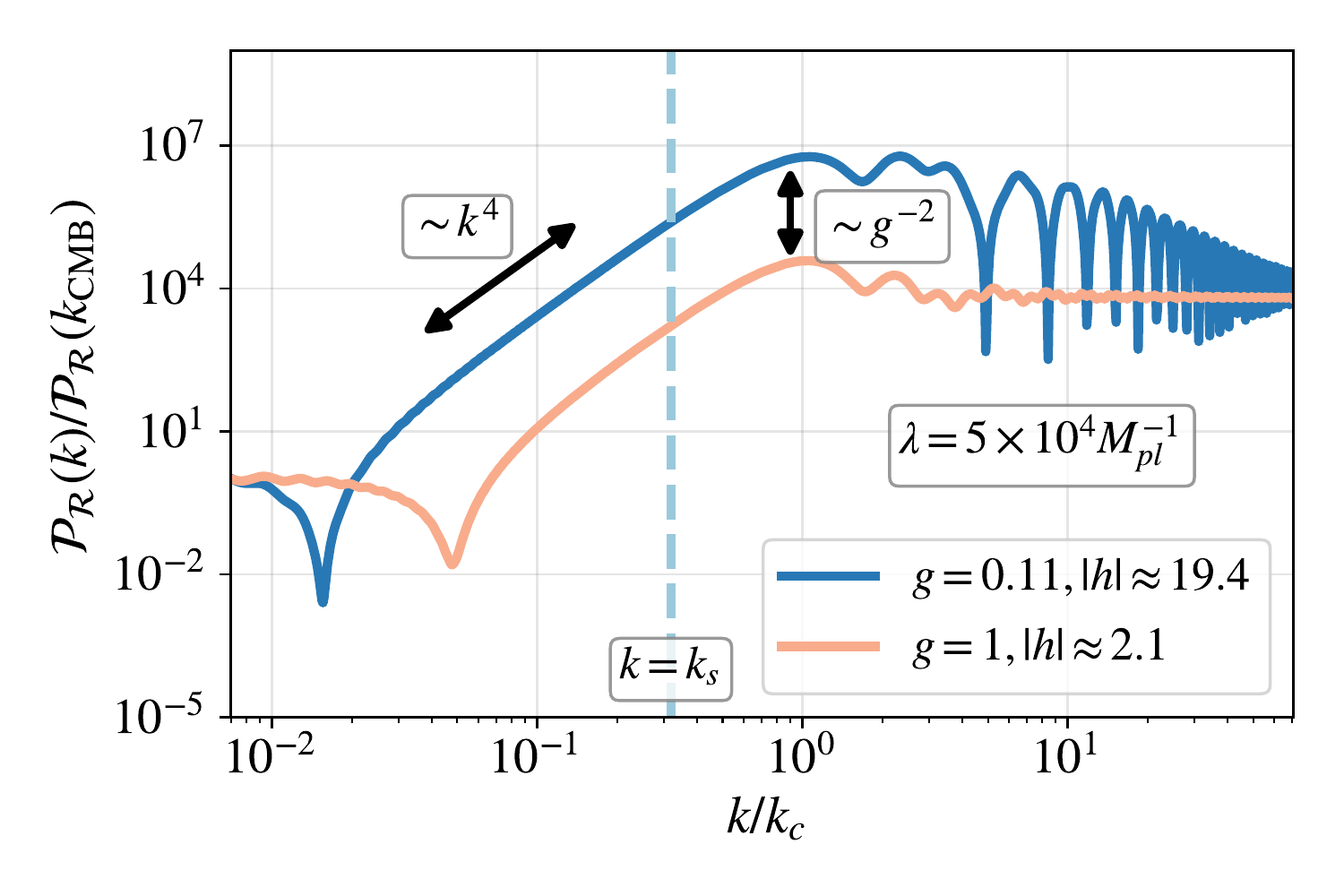}
     \includegraphics[width=0.45\textwidth]{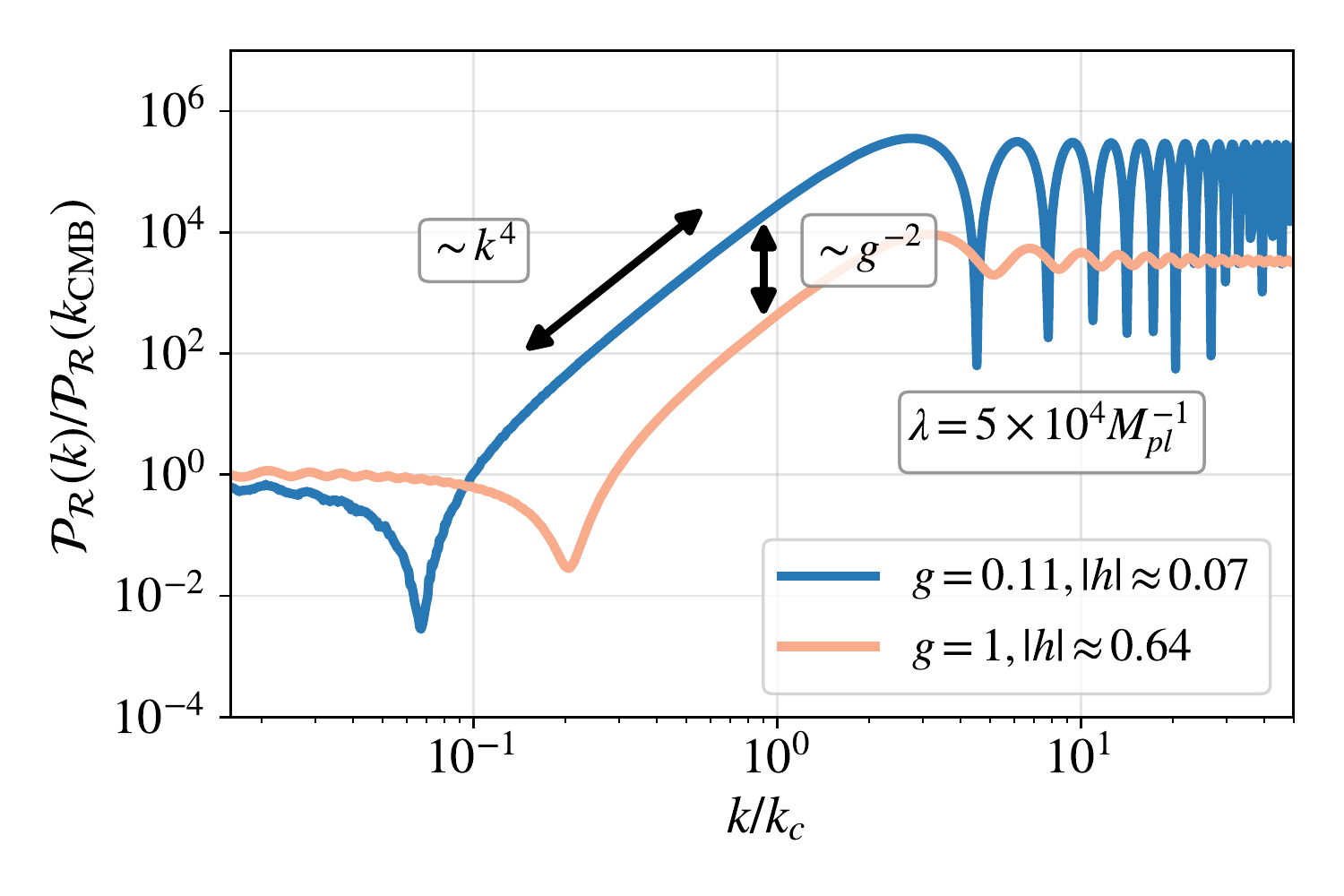}
    \caption{The enhancement of the power spectrum from both the USR phase and the upward step. The left panel shows the power spectrum of inflection-point model with USR phase which start at $\phi = \phi_s$ in figure \ref{fig:potential draft}. The right panel is the power spectrum without USR phase. The remaining parameters of the potential are the same except for $\Delta V = 5.85 \times 10^{-13}M^{4}_{\rm pl}$ to keep $g = 0.11$. The orange and blue curve corresponds to $g=1$ and $g=0.11$ respectively.  All the scaling properties match the behaviours displayed in Eq.~\eqref{Spectrum index} and \eqref{Power spectrum amplitude scaling} very well. }
    \label{fig:powerspectrum}
\end{figure}

To confirm our analytical estimates, we numerically computed the power spectrum by smoothing the step with the function, 
\begin{align}\label{Numerical step function}
\Theta(\phi) 
\equiv \frac{1}{2}\left( \tanh\left[\lambda (\phi-\phi_c)\right]  + 1 \right) ~,
\end{align}
where $\lambda$ controls the steepness of the step.
The left panel in figure~\ref{fig:powerspectrum} shows the result, where we chose a steep step  transition $\lambda = 5 \times 10^{4} M_{\rm pl}^{-1}$, with the energy scale $V_0 = 7 \times 10^{-10} M_{\rm pl}^{4}$. The resulting field velocity at the step is $\pi_c=-0.00233 M_{\rm pl}$. 
The values of the other parametes we chose are $\Delta V = 6.4 \times 10^
{-16}M^{4}_{\rm pl}$, $\epsilon_{S} = 2.551\times 10^
{-3}$, $\epsilon_{V} = 3.673\times 10^
{-7}$, $\eta_{V} = 7.143\times 10^
{-3}$ and the field range of USR $\Delta\phi_{USR} = \phi_{s} - \phi_{c} = 0.023M_{\rm pl}$.
This gives $\lambda|\pi_c|\approx 116$, which means the step is quite sharp.
Here we mention a difference from the analytical result. The analytical power spectrum oscillates with a constant amplitude as shown in \eqref{Short wavelength power spectrum}, while the numerical result shows damped oscillations, which is due to the finiteness of the step.
In contrast to the infinitely sharp approximation for the analytical computation, the width of the step determines the maximum wavenumber affected by the step, $k_\text{max}\sim \lambda\pi_c k_c$. The modes $k\gg k_\text{max}$ are not affected by the transition. 

To summarize, the analytical formula for the power spectrum for the potential~\eqref{parameterized potential} takes the form,
\begin{equation}
    \mathcal{P}_{\mathcal{R}}(k) \simeq \left\{
    \begin{aligned}
        & \frac{1}{2\epsilon_S} \(\frac{H}{2\pi}\)^2\,; & k \ll k_s ~, \\
        & \frac{1}{2\epsilon_V g^2} \(1- \frac{g^2 h}{6}\)^2 \(\frac{H}{2\pi}\)^2\,; 
        & k \lesssim k_c ~, \\
        & \frac{1}{2\epsilon_V} \frac{g^4 + 1 + (1-g^4)\cos(2k\tau_c)}{2g^2} \(\frac{H}{2\pi}\)^2\,; &k \gtrsim k_c ~,
    \end{aligned}
    \right.
\end{equation}
with the spectrum at $k\lesssim k_s$ having the power-law index $ n_s-1 \simeq 4$.

\subsubsection*{SR-SR Model}

The calculations for the SR-SR case are similar to those for the SR-USR-SR case discussed above. 
Before the step, the behaviour of the curvature perturbation is well approximated by the adiabatic mode function \eqref{eq:BD-vacuum mode function}. 
When the inflaton goes through the step, the negative frequency modes are excited as in the previous case.
Following the same technical details for the SR-USR-SR case, as given in App.\ref{Mode function matching}, the power spectrum for the parameter range of our interest, $g^2\ll |\eta_c| \ll 1$ and $|h|\sim \mathcal{O}(1)$, is given by
\begin{equation}
    \begin{aligned}
        \mathcal{P}_\mathcal{R}(k) &= \frac{H^2}{8\pi^2 \epsilon_k} |\alpha_k+\beta_k|^2\\ &\simeq
        \frac{H^2}{32\pi^2\epsilon_k  g^2  }\frac{   \left(\eta_c+2 k^2 \tau_c^2\right)^2+\eta_c^2 k^2 \tau_c^2}{k^6 \tau_c^6} \big[\sin (k \tau_c)-k \tau_c \cos (k\tau_c)\big]^2 ~.
    \end{aligned}
\end{equation}
Here the $\epsilon_k$ is the slow-roll parameter of the potential $\epsilon_V$ when $k$ modes crossing the Hubble horizon. The $k$-dependence of the power spectrum can be analyzed in three different scales. For the long wavelength modes with $k^2\tau_c^2\ll |\eta_c|\ll 1$, we have
\bea
    \mathcal{P}_\mathcal{R}(k)  \simeq
    \frac{H^2}{8\pi^2\epsilon_k } ~\frac{\eta_c^2}{36g^2} ,
\eea
while for the short wavelength modes $-k\tau_c\gg1$, we find
\be
    \mathcal{P}_\mathcal{R}(k)  \simeq
    \frac{H^2}{8\pi^2\epsilon_k  g^2  }  \qty[\cos (k\tau_c) -\frac{1}{k\tau_c}\sin(k\tau_c)]^2  \simeq  \frac{H^2}{8\pi^2\epsilon_k  g^2  } \cos^2(k\tau_c) .
\ee
For the intermediate frequencies with $|\eta_c|< k^2\tau_c^2< 1$, we obtain the $k^4$ growth behavior~\citep{Byrnes:2018txb},
\be
 \mathcal{P}_\mathcal{R}(k)  \simeq \frac{H^2}{8\pi^2\epsilon_k } ~\frac{k^4\tau_c^4}{9g^2}\,.
\ee
At $k \sim k_c$, the amplitude of the power spectrum is also proportional to $g^{-2}$.

To summarize, an approximate analytical formula for the power spectrum for the SR-SR transition model takes the form,
\begin{equation}
    \mathcal{P}_{\mathcal{R}}(k) \simeq \left\{
    \begin{aligned}
        & \frac{H}{8\pi^2 \epsilon_k} \frac{\eta^2}{36g^2}\,; &k^2 \tau_c^2 \ll |\eta_c| \ll 1 ~,\\
        & \frac{H^2}{8\pi^2 \epsilon_k} \frac{k^4 \tau_c^4}{9g^2}\,; &|\eta_c| < k^2\tau_c^2 < 1  ~, \\
        & \frac{H^2}{8\pi^2 \epsilon_k g^2}\cos^2\(k\tau_c\)\,;&k^2 \tau_c^2 \gg 1 ~, \\
    \end{aligned}
    \right.
\end{equation}
The right panel in figure~\ref{fig:powerspectrum} shows the corresponding numerical result. In this case, we have $\pi_c= -0.071M_{\rm pl}$, hence $\lambda|\pi_c|=710$.
Thus the step is much steeper than the SR-USR-SR case. This is the reason why
we do not see the decrease in the oscillation amplitude.

\subsection{Non-Gaussian imprints on PBH mass fraction}\label{PBH_mass_fraction}

With the above results at hand, we consider the implications of the non-Gaussian tail to the generation of PBHs. 
It is well-known that PBHs are formed from rare and large perturbations on the tail of the probability distribution.
To estimate the PBH abundance, it is customary to compute the fraction of the perturbation that turns into PBHs by
integrating the PDF of the density perturbation $P(\delta)$ above a certain critical value $\delta_c$ of order unity,
\begin{align}
    \beta_{\text{PBH}} = \int^{+\infty}_{\delta_c} \dd \delta\, P(\delta)\,.
\end{align}
We note that, for a Gaussian distribution with a small variance, $\left\langle{\delta^2}\right\rangle\ll1$, $\beta_\text{PBH}$ is extremely sensitive to the behaviour of the PDF at the tail. 
Although it is non-trivial in general to translate the critical value of the density perturbation to that of the curvature perturbation as the relation involves the Laplacian operator, for a peaked power spectrum, one may approximately evaluate $\beta_\text{PBH}$ by introducing a critical value $\calR_c$ corresponding to $\delta_c$ which is
also $\mathcal{O}(1)$ \cite{Musco:2020jjb}.
Thus we have
\begin{align}
    \beta_{\text{PBH}} = \int^{+\infty}_{\calR_c} \dd \calR \,P(\calR) \,,
\end{align}
where  $\calR_c=C\delta_c+\langle{\calR}\rangle$, where $C$ is a constant of order unity. 
As we are mainly interested in the primordial non-Gaussian tail, this simple approximation is good enough to show its effects on the PBH formation.


\begin{figure}[!htb]
    \centering
    \includegraphics[scale = 0.7]{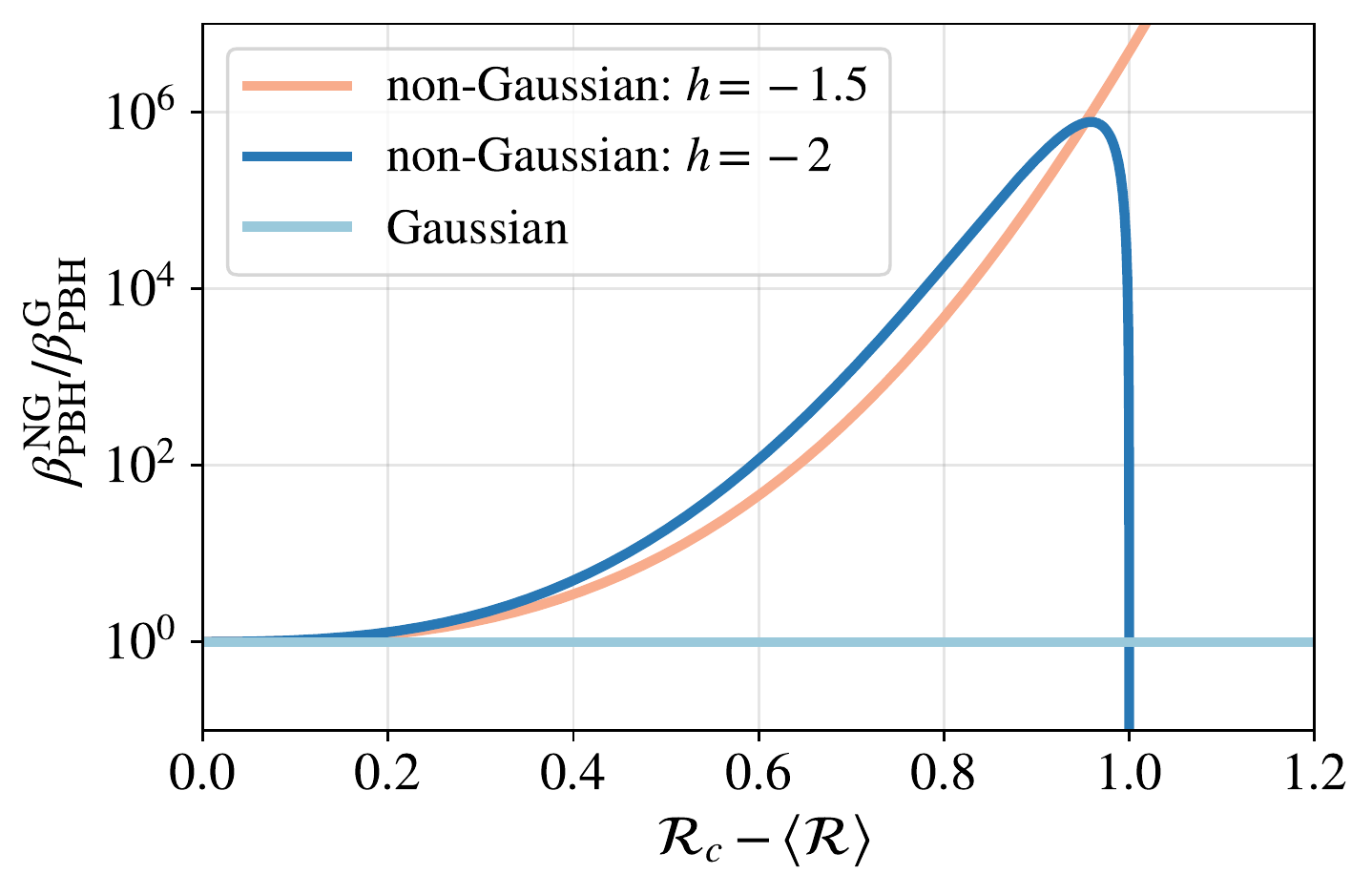}
    \caption{The ratio of PBH mass fractions from non-Gaussian and Gaussian tails.}
    \label{fig:beta}
\end{figure}

Using the PDF for $\calR$ in \eqref{pdfnon-G}, we obtain the mass fraction of PBHs at the time of formation,
\begin{align}
    \beta^{\text{NG}}_{\text{PBH}} &= \int_{\calR_c}^{2/\abs{h}}P(\calR)\dd \calR
    \nonumber\\
    &= \frac{\sqrt{2\pi\sigma^{2}_{\calR}}}{\Omega}\left[ \text{Erf}\left(\frac{1}{\abs{h}\sqrt{2\sigma^{2}_{\calR}}}\right) -  \text{Erf}\left(\frac{\calR_c\left(4-\abs{h}\calR_c\right)}{4\sqrt{2\sigma^{2}_{\calR}}}\right) \right]
    \Theta\left( \frac{2}{\abs{h}}-\calR_c\right) .\label{NG mass fraction in calR}
\end{align}
An intriguing fact is that there will be exactly no PBH if $2/|h|<\calR_c$. 
For comparison, the mass fraction for the Gaussian PDF \eqref{Gaussian} is given by
\begin{align}
    \beta^{\text{G}}_{\text{PBH}}  &= \int_{\calR_c}P(\calR_{\text{G}})\dd \calR_{\text{G}} = \frac{1}{2} \left[1-\text{Erf}\left(\frac{\calR_c}{\sqrt{2\sigma^{2}_{\calR}}}\right)\right] . \label{G mass fraction in calR}
\end{align}
As a demonstration of the enhancement of the PBH mass fraction, we plot the ratio of $\beta_\text{PBH}$ for the non-Gaussian case to the Gaussian case as a function of the critical value $\calR_C$ in figure~\ref{fig:beta}, for $\sigma^{2}_{\calR}=0.02$.
As the purpose of the paper is not to give a detailed analysis of the PBH formation, we ignore the ambiguities in the PBH formation critera discussed in the literature. 
As we can see, for $|h| \gtrsim1$, the non-Gaussian tail can enhance the PBH mass fraction 
by several orders of magnitude.


\subsection{Inflaton trapping as another seed for PBHs}\label{Subsec: trapping PBHs}

\begin{figure}[!htb]
    \centering
    \includegraphics[scale = 0.3]{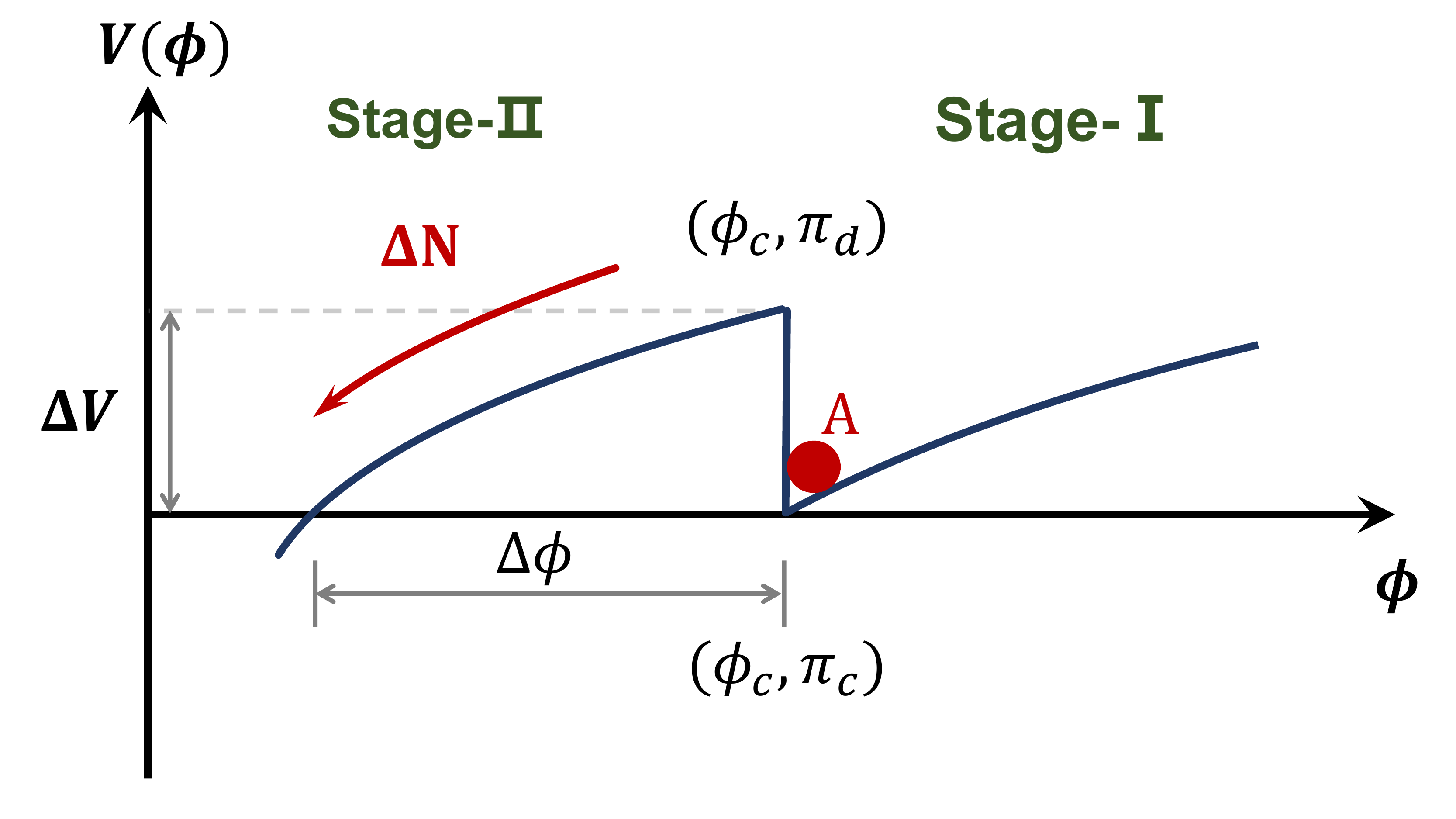}
    \caption{A sketch plot of inflaton trapping as the seeds of PBHs. When inflaton of region A is trapped in the bottom of the step, a bubble universe forms and begins to expand. Only when the potential energy of outside lower than $V(\phi_c^{+})$, the bubble universe start to collapse into a black hole.}
    \label{fig:trapped_PBH}
\end{figure}

Interestingly, in addition to the enhancement from the non-Gaussian tails,  there is another effect of upward-step models that can lead to the production of PBHs. Namely the trapping of the inflaton at the potential step \cite{Inomata:2021tpx}. When the inflaton arrives at $\phi_c$, the velocity fluctuations there may render some Hubble size regions of the universe trapped at $\phi_c$ due to an insufficient momentum to climb the step. 
Then, such a region of the universe surrounded by the region where the inflaton has successfully climbed up the step behaves like a true vacuum bubble in the sea of false vacuum.
Thus initially the bubble will expand exponentially. But as the inflaton in the exterior of the bubble rolls down the potential hill, the potential energy of the outer universe becomes smaller than that of the trapped region. Then the roles are reversed. That is, the trapped universe now behaves like false vacuum surrounded by true vacuum. 
Then, the bubble wall will be pushed toward the false vacuum side direction, and the
trapped region will become a black hole. This process is depicted in figure~\ref{fig:trapped_PBH}. 
In passing it is interesting to note that inside the bubble will still be expanding exponentially~\cite{Deng:2017uwc, Garriga:2015fdk}. Thus it gives rise to a wormhole-like spacetime geometry with two causally disconnected universes \cite{Sato:1981bf, Sato:1981gv}. 

We can estimate the mass of such PBHs as 
\begin{equation}
    M_{\rm PBH} = \rho \frac{4}{3}\pi R^3 \simeq \frac{4\pi M_{\rm pl}^2}{H} e^{3\Delta N} ~,
\end{equation}
where $\Delta N$ is the number of $e$-folds for the potential energy of the inflaton in the exterior of the bubble to become equal to that inside the bubble.
We can also estimate the probability of forming such PBHs. From Eq. \eqref{approximate solutioin of calR}, we can determine the trapping condition to be 
\begin{equation}
    \mathcal{R}_{\rm G} \geq \frac{1}{|h|} ~.
\end{equation}
Thus, the inflaton trapping probability, and hence the associated PBH formation probability 
is estimated as 
\begin{equation}
    \beta_{\rm PBH}^{\rm trap} = \frac{1}{2}\[1 + {\rm Erf} \(\frac{1}{\sqrt{2 \sigma_{\mathcal{R}}^2 } |h| }\)\] ~.
\end{equation}

Finally, let us mention the possibility that, due to quantum tunneling effects, the region trapped by the false vacuum may eventually tunnel to the true vacuum and thus destroying the PBH. According to \cite{Coleman:1977py}, the tunneling rate are actually exponentially suppressed by the Euclidean action $\Gamma\propto e^{-S_E}$. Typically, we would expect that, compared with the expansion rate of the universe, the tunneling probability is actually negligible, i.e. $\Gamma/H^4 \ll 1$. The detailed computation here is actually model-dependent and we refer to readers to a specific example mentioned in \cite{Inomata:2021tpx}.

\section{Conclusions and outlook}

In the study of the primordial curvature perturbation, the tail of the probability distribution is a wonderland where the conventional perturbative approaches break down, and large but rare fluctuations can lead to plentiful phenomenological consequences such as the abundant PBH formation. Interestingly, deviations from the Gaussian distribution of the tail may not be necessarily related to the lower-order moments computed in the perturbative regime. Therefore, non-Gaussian tails may provide us with a novel and crucial window for probing non-perturbative effects during inflation.

In this paper, we argued that the tail of the probability distribution of the curvature perturbation can be highly non-Gaussian even if the perturbative non-Gaussianity remains small. In particular, we constructed a specific model of single-field inflation with an upward step in the potential that supports our argument. We performed a detailed analysis of the background evolution with an upward-step transition, with or without an USR stage, and identified the important role played by off-attractor trajectories in the phase space. Then through a perturbative computation, we find that the local non-Gaussianity in these models depends on the details of the upward step transition, which can be either large or small. 
In particular, the local $\fnl$ parameter can become much bigger than the one for the original USR inflation $\fnl=5/2$.

Then focusing on the non-Gaussian tail from the step transition, which is the key part of this work, we derived a non-perturbative $\delta N$ formula, and identified a nonlinear mapping from the Gaussian inflaton fluctuation to the curvature perturbation. This relation leads to highly nontrivial tail behaviour of the non-Gaussian PDF of the curvature perturbation. Intriguingly, we found that the tail of the distribution is significantly suppressed in the case of large $\fnl$, while it is exponentially enhanced in the case of small $\fnl$. This result,
in apparent contradiction with the naive speculation from the perturbative analysis,
is due to the non-perturbative effects caused by the upward step in both the 
USR-SR and SR-SR transitions.

Lastly, we studied the implications of our results to the formation of PBHs. For this purpose, we explicitly computed the curvature perturbation spectrum for an inflection-point potential model in which inflation begins with a SR stage, goes through an USR stage, and makes a transition to another SR stage with an upward step in the potential.
Our analysis shows that due to the highly non-perturbative non-Gaussian tail, the mass fraction of PBHs can be boosted by several orders of magnitudes. In addition, in such a kind of single-field inflation models with an upward step, we showed that there exists another possibility of producing PBHs by trapping the inflaton at the bottom of the step. Namely, some regions of the space where the inflaton failed to climb up the upward step would eventually collapse to black holes.

The above results can also inspire  new lines of research for novel phenomenologies of non-Gaussian tails.
Here we close by listing several possible  directions for future investigations.
First of all, it would be interesting to further explore various  non-Gaussian tails in other scenarios with non-perturbative effects.
As demonstrated in our study, the fully nonlinear mapping between $\calR$ and $\delta \phi$ plays a key role for the tail of the PDF. Meanwhile, the stochastic effects during inflation may also lead to nontrivial non-Gaussian tails
\cite{Ezquiaga:2019ftu, Figueroa:2020jkf, Pattison:2021oen, Achucarro:2021pdh, Ahmadi:2022lsm}. 
Thus it is encouraging to extend our current analysis to incorporate more general considerations.
Secondly, for the studies on PBHs, there are various mechanisms in the literature for their formation, such as resonance effects during inflation \cite{Cai:2018tuh, Cai:2019jah, Chen:2019zza, Chen:2020uhe, Zhou:2020kkf, Cai:2021wzd, Cai:2021yvq, Peng:2021zon} and multifield dynamics \cite{Pi:2021dft, Hooshangi:2022lao}. Richer behaviours of non-Gaussian tails may be expected in these scenarios, which deserve a closer look.
At last, in addition to PBHs, it would also be crucial to study other phenomenological implications of the non-Gaussian tail. One  possibility is  the scalar-induced Gravitational Waves during inflation.
Normally their amplitudes are also amplified when there are enhanced curvature perturbations for generating PBHs. Then the presence of a highly non-Gaussian tail could lead to distinct signals in these induced tensor modes. These open questions deserve a closer look in future research.


\


\paragraph*{Acknowledgements}
We are grateful to Chao Chen, Xingang Chen, Keisuke Inomata, Bichu Li, Chunshan Lin, Mohammad Hossein Namjoo, Bo Wang and Yi Wang for discussions.
DGW thanks the University of Science and Technology of China for hospitality where this work was initiated. 
YFC and XHM are supported in part by the National Key R\&D Program of China (2021YFC2203100), the NSFC (11961131007, 12261131497), by the Fundamental Research Funds for Central Universities, by the CSC Innovation Talent Funds, by the CAS project for young scientists in basic research (YSBR-006), by the USTC Fellowship for International Cooperation, and by the USTC Research Funds of the Double First-Class Initiative.
MS is supported in part by JSPS KAKENHI grants (19H01895, 20H04727, 20H05853).
DGW is supported by the Netherlands Organisation for Scientific Research (NWO) through a Vidi grant with Project No. 680-47-535, and a Rubicon Postdoctoral Fellowship.
ZZ is supported in part by the scholarship at Princeton.
We acknowledge the use of computing facilities of Kavli IPMU, as well as the clusters {\it LINDA} and {\it JUDY} of the particle cosmology group at USTC.

\appendix

\section{The $\pi$-dependence at the end of the initial SR stage}\label{pi-dependence}\label{App: pi-dependence before the step}

In this appendix, we examine the $\pi$-dependence of phase-space trajectories on a SR potential with detailed computations.
In particular, we distinguish between the converging and the non-converging regimes of these trajectories, and highlight the importance of the off-attractor behaviour around the end of the SR phase.

To explain how the expansion history of the Universe depends on the initial condition $(\phi_i,\pi_i)$ at the slow-roll phase before the step, we need to look at the full equation of motion of the inflaton $\phi$ without slow-roll approximations.
Consider the inflaton starts rolling with the initial condition $(\phi_i,\pi_i)$ on a SR potential parameterized by
\begin{equation}
V(\phi) = V_0 \left[ 1+ \sqrt{2\epsilon_{\rom{1}}}\left(\phi -\phi_c \right) + \frac{1}{2}\eta_{\rom{1}}\left(\phi -\phi_c \right)^2\right]~, ~~~~{\rm for }~ \phi>\phi_c~. 
\end{equation}
For a small range of field excursion before the end of this SR stage at $\phi_c$, we have $3H^2\simeq V_0$, and the full equation of motion,
\begin{align}
    \frac{d^{2} \phi}{d n^{2}}+3 \frac{d \phi}{dn}+3
    \sqrt{2 \epsilon_{\rom{1}}}+3 \eta_{\rom{1}}\left(\phi-\phi_{c}\right) = 0 ~,~~~~{\rm for}~~ \phi > \phi_c ~,
    \label{Background equation of inflaton in phase1}
\end{align}
which has the general solution,
\begin{equation}
    \begin{aligned}
        \phi(n) =& C_1 e^{\frac{1}{2}(s-3) (n-n_i)}
        - C_2 e^{-\frac{1}{2}(s+3) (n - n_i)}
        +\frac{12 \sqrt{2\epsilon_{I}}}{s^{2}-9}+\phi_{c} ~.
    \end{aligned}\label{full phi solution in stage 1}
\end{equation}
The parameter $s$ is defined in the same way as in \eqref{hs}, $s \equiv \sqrt{9 - 12\eta_{\rom{1}}} \simeq 3 - 2\eta_{\rom{1}}$.
The dependence of initial condition $(\phi_i,\pi_i)$ is contained in two coefficients $C_1$ and $C_2$, with
\begin{equation}
    \begin{aligned}
        C_1 &= \[-\frac{6\sqrt{2\epsilon_{I}}}{s(s-3)}+\frac{1}{2s}\( 2\pi_i +(s+3)(\phi_i-\phi_c)\)\] \simeq \frac{1}{6}\qty[ \pi_i + 6(\phi_i-\phi_c) + 
        \frac{12\sqrt{2\epsilon_{I}}}{\eta_{\rom{1}}} ]~,\\
        C_2 &= \[\frac{6\sqrt{2\epsilon_{I}}}{s(s+3)}+\frac{1}{2s}\( 2\pi_i - (s-3)(\phi_i-\phi_c)\)\] \simeq \frac{1}{6}\qty[ \pi_i + \eta_{\rom{1}}(\phi_i-\phi_c) +  \sqrt{2\epsilon_{I}} ] ~.
    \end{aligned}
\end{equation}
Meanwhile, $\pi(n)$ can be computed by taking the derivative of \eqref{full phi solution in stage 1}
\begin{equation}
    \begin{aligned}
        \pi(n) = \frac{s-3}{2}C_1 e^{\frac{1}{2}(s-3) (n-n_i)} 
        + 
        \frac{s+3}{2}C_2 e^{-\frac{1}{2}(s+3) (n - n_i)} ~.
    \end{aligned}\label{full pi solution in stage 1}
\end{equation}

The solutions \eqref{full phi solution in stage 1} and  \eqref{full pi solution in stage 1} give us all the possible phase-space trajectories without slow-roll approximations.
There are two parts in these solutions: the one with the coefficient $C_1$ is a nearly constant piece; while the one with $C_2$ is rapidly decaying one. 
When the initial condition $(\phi_i,\pi_i)$ is on the slow-roll attractor, one can simply check that $C_2\simeq0$ and the $C_1$ piece of the solution reproduces the slow-roll results.

Here we would like to consider in general how   trajectories with arbitrary initial conditions $(\phi_i,\pi_i)$ converge into the slow-roll attractor, and identify how the number of e-folds depends on the non-slow-roll initial conditions. 
We first perform the analytical computation in two different regimes of these trajectories:

\begin{itemize}
\item {\it The converging regime, for $n-n_i\gtrsim \mathcal{O}(1)$}. By using  \eqref{full phi solution in stage 1} and \eqref{full pi solution in stage 1}, and eliminating the $C_1$ terms, we find these trajectories satisfy the relation,  
\begin{align}
    \phi(n) = -\frac{\pi(n)}{\eta_{\rom{1}}} - \frac{-\sqrt{2 \epsilon_{\rom{1}}}}{\eta_{\rom{1}}}+ \phi_c 
    +e^{-(3-\eta_{\rom{1}})(n-n_i)}\qty[(\phi_i - \phi_c) + \frac{\pi_i}{\eta_{\rom{1}}} - \frac{\sqrt{2 \epsilon_{\rom{1}}}}{\eta_{\rom{1}}}]~.
    \label{trajectories}
\end{align}
Thus, when the duration is long enough $n-n_i\gtrsim \mathcal{O}(1)$, the dependence on $\pi_i$ from the last term is exponentially suppressed. This corresponds to the situation where the trajectories have converged into the slow-roll attractor.
If we neglect this last term in \eqref{trajectories}, it simply reduces to the result with slow-roll conditions,
\begin{align}
    \pi(n) \simeq -\sqrt{2 \epsilon_{\rom{1}}} + \eta_{\rom{1}}\qty[\phi_c - \phi(n)]  ~.
    \label{base trajectory}
\end{align}
If we assume the SR stage ends in this regime, i.e. $\phi(n_c) = \phi_c$ with $n_c-n_i\gtrsim \mathcal{O}(1)$, the number of e-folds $N_{\rom{1}}=n_c-n_i$ can be solved from \eqref{full pi solution in stage 1} analytically by dropping the $C_2$ term 
\begin{align}\label{N1 with slow-roll approx}
    N_{\rom{1}} = n_c - n_i =-\frac{1}{\eta_{\rom{1}}}\log \qty[\frac{\pi(n_c)}{\eta_{\rom{1}} C_1(\phi_i,\pi_i)}]\simeq \frac{1}{\eta_{\rom{1}}}\log\left[ 1 + \frac{\eta_{\rom{1}}}{\sqrt{2\epsilon_{\rom{1}}}}\left(\phi_i - \phi_c\right) \right] ~. 
\end{align}

\item {\it The non-converging regime, for $n-n_i \ll \mathcal{O}(1)$}. Now we eliminate the $C_2$ terms in \eqref{full phi solution in stage 1} and \eqref{full pi solution in stage 1}, and then find
\begin{equation} \label{off-traj}
 \frac{s+3}{2} \phi(n) +\pi(n) = C_1 s e^{\frac{1}{2}(s-3) (n-n_i)} +\frac{6 \sqrt{2\epsilon_{I}}}{s-3}+\frac{s+3}{2}\phi_{c}~.
\end{equation}
This regime corresponds to the initial evolution with non-slow-roll initial conditions before the attractor is reached. 
Its duration is expected to be very short, and from \eqref{off-traj} we find approximately
\be
 3\phi(n) + \pi(n) \simeq 3\phi_i + \pi_i~,
\ee
which has the non-attractor dynamics. We are interested in trajectories that are still in this non-converging regime when the inflaton reaches the end of the SR potential at $\phi_c$. In such a situation, we find the number of e-folds $N_I$ is given by
\be
 N_I =\frac{1}{3} \log\[\frac{\pi_i}{\pi_c}\]= \frac{1}{3} \log\[ \frac{\pi_i}{\pi_i+3(\phi_i-\phi_c)}\].
\ee
These are the off-attractor trajectories at the end of the initial SR stage  discussed in Section \ref{sec:USR_SR} and \ref{sec:SR to SR}.
\end{itemize}

\begin{figure}[!htb]
    \centering
    \includegraphics[scale = 0.8]{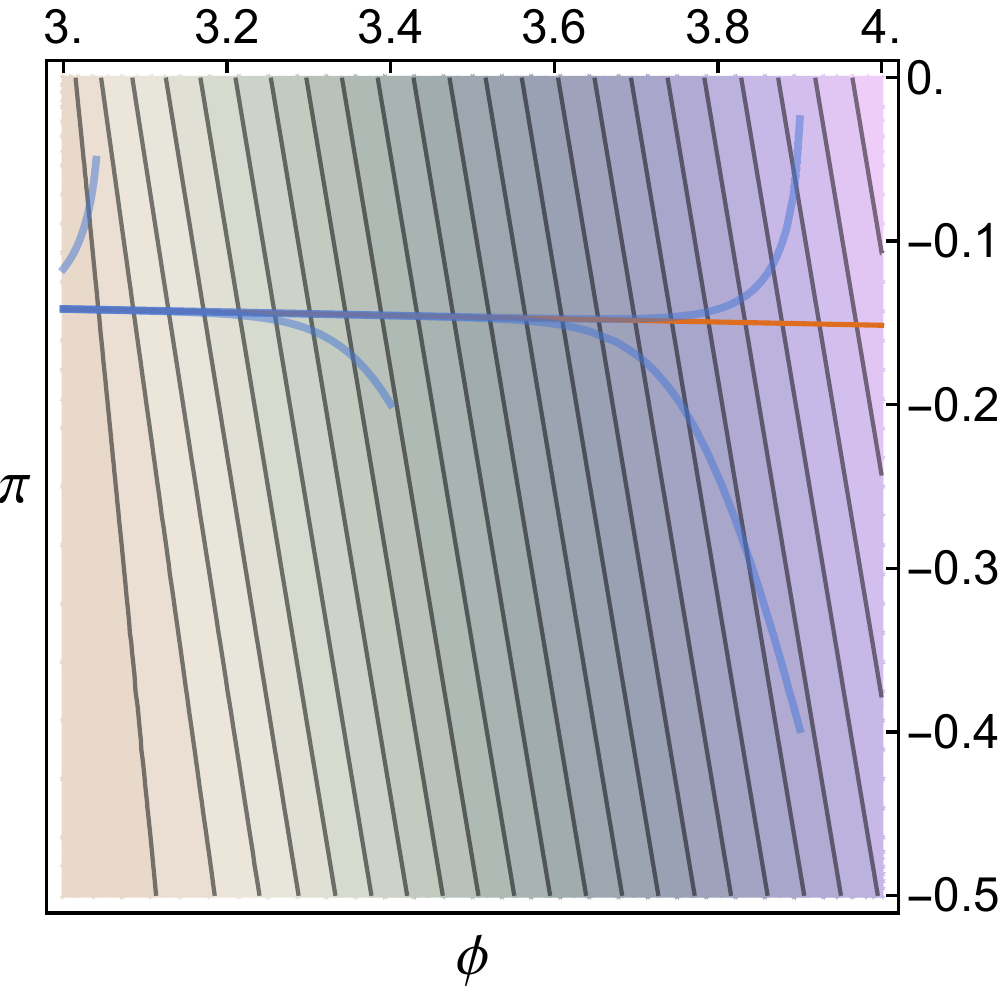}
    \caption{The contours of $N_{\rom{1}}$ in the phase space which demonstrates its dependence on the initial conditions $(\phi_i,\pi_i)$. In this numerical result, we set the end of the initial SR stage is $\phi_c=3$. As shown in \eqref{N1 with slow-roll approx}, there is mild $\pi_i$-dependence in $N_{\rom{1}}$   in the  $N_{\rom{1}}\gtrsim\mathcal{O}(1)$ region. Red line is the slow-roll attractor given by \eqref{base trajectory} with $\eta_{\rom{1}} = 0.01$, $\epsilon_{\rom{1}} = 0.01$. The blue line shows how trajectories converge to slow-roll attractor with different initial conditions. }
    \label{fig:contour of E-folds}
\end{figure}

In general, there is no explicit formula for the number of e-folds $N_I$ beyond the two limiting regimes above. We solve \eqref{full phi solution in stage 1} and \eqref{full pi solution in stage 1} numerically and demonstrate how $N_I$ depends on initial conditions $(\phi_i,\pi_i)$ in figure \ref{fig:contour of E-folds}.

\section{Mode function matching}\label{Mode function matching}

\subsection{$\delta\phi$ gauge}\label{deltaphi gauge}

In the analytical model of the upward step transition studied in Section 2, the evolution of the mode function $\delta\phi$ can be determined straightforwardly. In the conformal time, the standard perturbation equation takes the form as follows,
\begin{equation}\label{deltaphi MS}
 \delta\phi''+2\mathcal{H}\delta\phi' + (k^2 + a^2 V_{,\phi\phi})\delta\phi = 0 ~.
\end{equation}
Both in the USR and SR stages, $V_{,\phi\phi}$ is expected to be small enough to be negligible. As a result, this leads to the field perturbation mode function as
\begin{equation}
    \delta\phi_k(\tau) = \frac{H}{\sqrt{2 k^3}}(1+ik\tau)e^{-ik\tau} ~,
\end{equation}
and its conformal derivative takes
\begin{equation}
    \delta\phi_k'(\tau) = \frac{H}{\sqrt{2k^3}}k^2\tau e^{-ik\tau} ~.
\end{equation}

\begin{figure}[H]
    \centering
    \includegraphics[scale = 0.4]{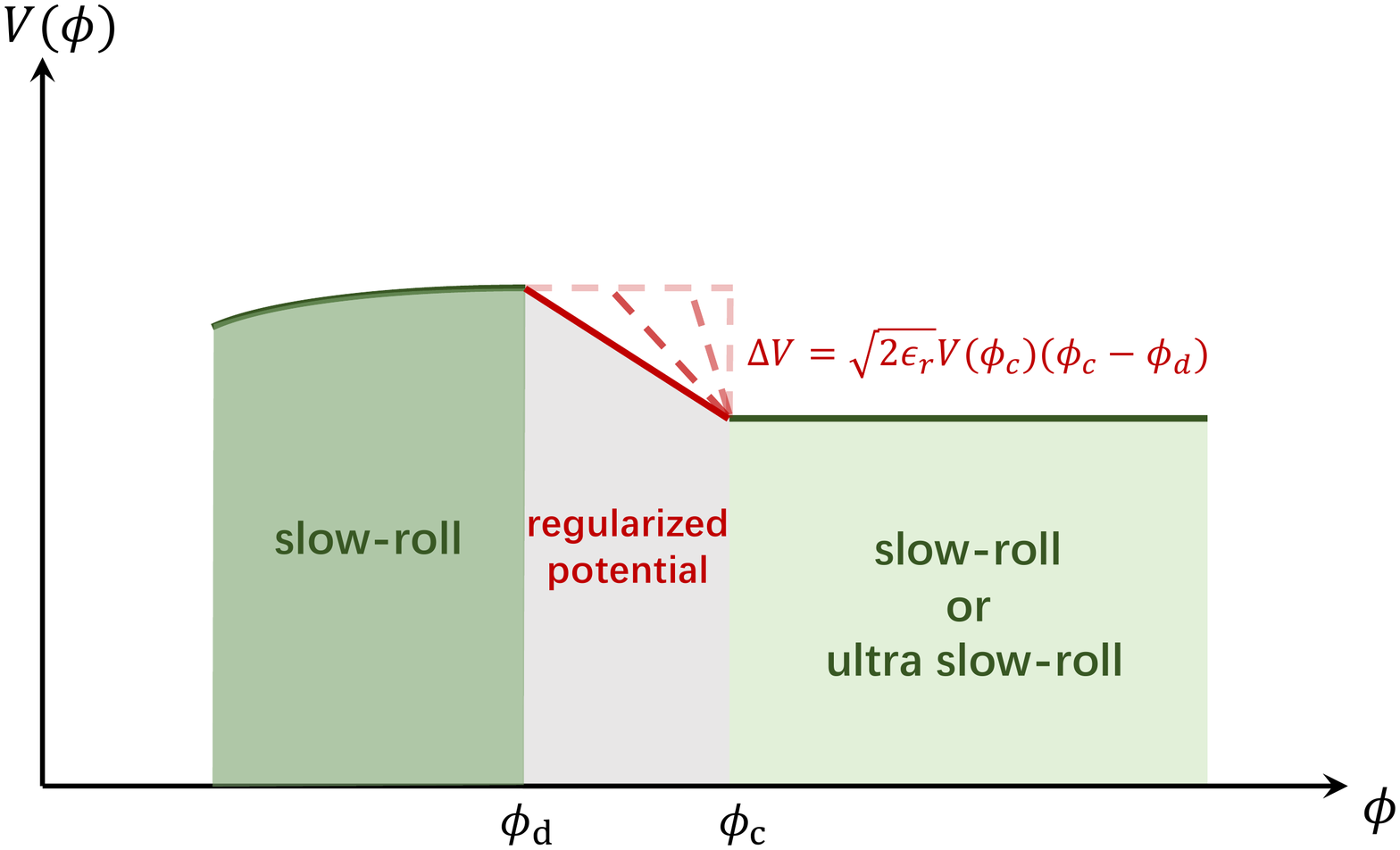}
    \caption{A sketch figure showed our regularization method. A linear potential connects $\phi_c$ and $\phi_d$ instead of a step. The slope of potential in this stage depends on $\phi_c - \phi_d$, with a fixed $\Delta V$ in a concrete case. Therefore, the result calculated in this case depends on $\phi_c - \phi_d$. When choosing step limit, means to take $\phi_c - \phi_d$ go to $0$, correct solution for the step case get. Dashed lines of different transparencies in the figure illustrate this process.}
    \label{regularized_potential}
\end{figure}

However, here $V_{,\phi\phi}$ is not well-defined at the transition stage. We need to look for a certain regularization procedure to tackle this problem. In our case, we choose a linear potential to connect $\phi_c$ and $\phi_d$ as figure~\ref{regularized_potential} shows. More concretely, let us suppose that the potential connects $\phi_c$ and $\phi_d$

\begin{equation}
    V(\phi) = V(\phi_c)-\sqrt{2\epsilon_r}V(\phi_c)(\phi-\phi_c) ~,
\end{equation}
which leads to the background equation
\begin{equation}
    \frac{d^2 \phi}{dN^2}+3\frac{d\phi}{dN}-3\sqrt{2\epsilon_r}\simeq 0 ~.
\end{equation}
Without losing generality, we set $N=0$ at $\phi_c$. Then, for given $|\pi_c|>|\pi_d|>0$, we have the following solution
\begin{equation}\label{BGEqCD}
    \phi = \frac{1}{3} e^{-3 N}\left(\sqrt{2 \epsilon_r}- e^{3 N} \sqrt{2 \epsilon_r}+3 \mathrm{e}^{3 N} N \sqrt{2 \epsilon_r}-\pi_c +\mathrm{e}^{3 N} \pi_c+3 \mathrm{e}^{3 N} \phi_c \right) ~,
\end{equation}
and the duration between $\phi_c$ and $\phi_d$ is given by
\begin{equation}
    N_d = \frac{1}{3}\log \left[ \frac{2\epsilon_r - \sqrt{2\epsilon_r}(\pi_c - \pi_d) - \pi_c \pi_d}{2\epsilon_r -\pi_d^2}\right] ~.
\end{equation}
After computing the mode function, we take $\phi_c\rightarrow \phi_d$ in the end. In this case, we can write $V_{,\phi\phi}$ as
\begin{equation}
    V_{,\phi\phi} \simeq \Big[V'(\phi_{d_+})-V'(\phi_{d_-})\Big]\delta(\phi-\phi_d) + \Big[V'(\phi_{c_+})-V'(\phi_{c_-})\Big]\delta(\phi-\phi_c) ~.
\end{equation}

We can write Eq.\eqref{deltaphi MS} in the canonical form,
\begin{equation}
    \delta\Phi_k'' + \Big[k^2-\frac{a''}{a} + a^2 V_{,\phi\phi}\Big]\delta\Phi_k = 0 ~,
\end{equation}
where $\delta\Phi=a\delta\phi$. Integrating this equation from $\tau_{c_-}$ to $\tau_{c_+}$, $\tau_{d_-}$ to $\tau_{d_+}$ and making use of the formula $\delta(\phi(\tau)-\phi_d) = \delta(\tau-\tau_d)/|\phi'(\tau_d)|$, $\delta(\phi(\tau)-\phi_c) = \delta(\tau-\tau_c)/|\phi'(\tau_c)|$, we get the matching condition,
\begin{equation}
    \delta\phi(\tau_{c_-}) = \delta\phi(\tau_{c_+})~, \quad \delta\phi'(\tau_{c_-}) + a^2\frac{V'(\phi_{c_-})}{|\phi'(\tau_{c_-})|} =  \delta\phi'(\tau_{c_+}) + a^2\frac{V'(\phi_{c_+})}{|\phi'(\tau_{c_+})|}~.
\end{equation}
and
\begin{equation}
    \delta\phi(\tau_{d_-}) = \delta\phi(\tau_{d_+})~, \quad \delta\phi'(\tau_{d_-}) + a^2\frac{V'(\phi_{d_-})}{|\phi'(\tau_{d_-})|} =  \delta\phi'(\tau_{d_+}) + a^2\frac{V'(\phi_{d_+})}{|\phi'(\tau_{d_+})|}~.
\end{equation}

This gives the following solutions of curvature perturbations
\begin{align}
    \delta\phi_{k}(\tau)&=\alpha^{r}_{k} \frac{H}{\sqrt{2 k^{3}}}(1+i k \tau) e^{-i k \tau}+\beta^{r}_{k} \frac{H}{\sqrt{2 k^{3}}}(1-i k \tau) e^{i k \tau} ~; \tau_c < \tau < \tau_d\,,\\
    \delta\phi_{k}(\tau)&=\alpha_{k} \frac{H}{\sqrt{2 k^{3}}}(1+i k \tau) e^{-i k \tau}+\beta_{k} \frac{H}{\sqrt{2 k^{3}}}(1-i k \tau) e^{i k \tau}~; \tau > \tau_d\,,
\end{align}
and their derivatives
\begin{align}
&\delta\phi_{k}^{\prime}(\tau)=\alpha^{r}_{k} \frac{H}{\sqrt{2 k^{3}}} k^{2} \tau e^{-i k \tau} + \beta^{r}_{k} \frac{H}{\sqrt{2 k^3}} k^{2} \tau e^{i k \tau} ~; \tau_c < \tau < \tau_d\,,
 \\
&\delta\phi_{k}^{\prime}(\tau)=\alpha_{k}\frac{H}{\sqrt{2 k^{3}}} k^{2} \tau e^{-i k \tau} + \beta_{k} \frac{H}{\sqrt{2 k^{3}}} k^{2} \tau e^{i k \tau} ~; \tau > \tau_d\,,
\end{align}
where
\begin{align}
\alpha^{r}_{k}=1-\frac{i h_c\left(1+k^{2} \tau_c^{2}\right)}{4 k^{3} \tau_c^{3}}, \quad \beta^{r}_k = -\frac{i e^{-2i k \tau_c}h_c (-i+k\tau_c)^2}{4k^3\tau_c^3} ~,
\label{C Match}
\end{align}
and
\begin{equation} \label{D Match}
    \begin{aligned}
        \begin{aligned}
            \alpha_k = &\frac{1}{16 k^{6} \tau_c^{3} \tau_d^{3}} e^{-2 i k \tau_c}\Big[-e^{2 i k \tau_d} (-6 + h_c - \eta_c)(h+h_d)(-i+k \tau_c)^{2}(i+k \tau_d)^{2} \\
            & + e^{2 i k \tau_c} \left((-6 + h_c - \eta_c)(1+ k^{2} \tau_c^{2}) + 4 i k^{3} \tau_c^{3}\right)\left(h+h_d+(h+h_d) k^{2} \tau_d^{2}-4 i k^{3} \tau_d^{3}\right)\Big]
        \end{aligned}\\
        \begin{aligned}
            \beta_k = &\frac{1}{16 k^{6} \tau_c^{3} \tau_d^{3}} e^{-2i k (\tau_c+\tau_d)}\Big[e^{2ik\tau_c} (h+h_d) ((-6 + h_c - \eta_c)(1 + k^2\tau_c^2) + 4 i k^3 \tau_c^3)(-i+k\tau_d)^2 \\
            & - e^{2 i k \tau_d}(-6 + h_c - \eta_c)(-i+k \tau_c)^2 (h+h_d +(h+h_d)k^2 \tau_d^2 + 4 i k^3 \tau_d^3) \Big] ~.
        \end{aligned}
    \end{aligned}
\end{equation}
For the convenience, here we introduce $h_c=6\sqrt{2\epsilon_r}/\pi_c$ and $h_d = 6\sqrt{2\epsilon_r}/\pi_d$.
$\eta_c \equiv \dot{\epsilon_c}/H\epsilon_c$ is the second slow roll parameter at the trasition point $\phi = \phi_c$. For USR to SR transition(the model in section \ref{sec:USR_SR}) $\eta_c = -6$, and for SR to SR transition(the model in section \ref{sec:SR to SR}) $\abs{\eta_c} \ll 1$. In our regularization method, steepness of the upward step was described by $\epsilon_r$.

It is also easy to find that if we take the limit $\phi_c\rightarrow \phi_d$, it is equivalent to take $\epsilon_r\rightarrow +\infty$ while keeping $\pi_c$ and $\pi_d$ fixed. In this limit, the matching conditions yield
\begin{equation}\label{Limit D Match}
    \begin{aligned}
        \alpha_k &= \frac{i \left(g^2 (h+6) + \eta_c\right)}{4 g k^3 \tau_c^3}+\frac{i \left(g^2
        (h+4)+2+\eta_c \right)}{4 g k \tau_c}+\frac{g^2+1}{2 g} ~, \\
        \beta_k &= -\frac{i \left(g^2 (h+6)+ \eta_c\right) e^{-2 i k \tau_c}}{4 g k^3
        \tau_c^3} +\frac{\left(g^2 (h+6)+ \eta_c \right) e^{-2 i k \tau_c}}{2 g k^2 \tau_c^2} \\ & \quad +\frac{i \left(g^2 (h+8)+ \eta_c -2 \right) e^{-2 i
        k \tau_c}}{4 g k \tau_c}-\frac{\left(g^2-1\right) e^{-2
        i k \tau_c}}{2 g} ~.
    \end{aligned}
\end{equation}

Just as what we did in Sec.\ref{matching}, we introduce a reference mode $k_c$, which satisfies $k_c \tau_c = -1$. For the long wavelength modes $k < k_c$, to the leading order, there are
\begin{equation}
    \begin{aligned}
        \alpha_k &= \frac{i \left(g^2 (h+6) + \eta_c\right)}{4 g k^3 \tau_c^3} ~, \quad k < k_c ~, \\
        \beta_k &= -\frac{i \left(g^2 (h+6) + \eta_c\right) e^{-2 i k \tau_c}}{4 g k^3
        \tau_c^3} ~, \quad k < k_c ~.
    \end{aligned}
\end{equation}
Meanwhile, for the short wavelength modes $k > k_c$, one gets
\begin{equation}
    \begin{aligned}
        \alpha_k &\simeq \frac{g^2+1}{2 g} ~,\quad k > k_c~, \\
        \beta_k &\simeq -\frac{\left(g^2-1\right) e^{-2
        i k \tau_c}}{2 g} ~, \quad k > k_c ~.
    \end{aligned}
\end{equation}

\subsection{$\calR$ gauge}\label{R gauge}
In this appendix, we provide the details of calculating the mode function in $\mathcal{R}$ gauge as a double check of the results in Appendix \ref{deltaphi gauge}. First of all, for the primordial perturbation $\calR$, the Mukhanov-Sasaki equation takes the form
\begin{equation}
    u_k''+\Big(k^2-\frac{z''}{z}\Big)u_k=0
\end{equation}
where $u_k\equiv z\calR_k$ and $z\equiv a\sqrt{2\epsilon}$. Following the standard treatment, the effective mass term can be written as
\begin{equation}
    \frac{z''}{z}=(aH)^2\Big[2-\epsilon + \frac{3}{2}\eta -\frac{1}{2}\epsilon \eta +\frac{1}{4}\eta^2 + \frac{\dot\eta}{2H}\Big] ~,
\end{equation}
In the USR stage, $\eta=-6$ and thus $z''/z\simeq 2/\tau^2$. As a result, this leads to the curvature perturbation mode function as
\begin{equation}
    \calR_k(\tau)=\frac{H}{\sqrt{4\epsilon k^3}}(1+ik\tau)e^{-ik\tau} ~,
\end{equation}
and its conformal time derivative
\begin{equation}
    \calR_{k}^{\prime}(\tau)=\frac{H}{\sqrt{4 \epsilon k^{3}}} k^{2} \tau e^{-i k \tau}-\frac{\eta}{2} a H \frac{H}{\sqrt{4 \epsilon k^{3}}}(1+i k \tau) e^{-i k \tau}
\end{equation}
which shares the same form with the lowest order slow-roll approximation. Meanwhile, with the help of Eq.~\eqref{background equation in slow-roll} we find that even though $\eta$ and $\dot \eta$ varies dramatically during the transition, $z''/z = (2-3\eta_V)/\tau^2$ is always satisfied, except at $\phi_c$ and $\phi_d$.

To tackle the problem of ill-defined $\eta$ during the transition stage, we need to perform the similar procedure as mentioned in the previous section. Due to the sudden change of $\eta$ at the transition at $\phi_c$ and $\phi_d$, the matching conditions force the mode function and its first derivative to be continuous, i.e. $\calR(\tau_{c_-})=\calR(\tau_{c_+})$, $\calR(\tau_{d_-})=\calR(\tau_{d_+})$ and $\calR'(\tau_{c_-})=\calR'(\tau_{c_+})$, $\calR'(\tau_{d_-})=\calR'(\tau_{d_+})$. This gives the following behaviour of curvature perturbations
\begin{align}
    \calR_{k}(\tau)&=\alpha^{r}_{k} \frac{H}{\sqrt{4 \epsilon k^{3}}}(1+i k \tau) e^{-i k \tau}+\beta^{r}_{k} \frac{H}{\sqrt{4 \epsilon k^{3}}}(1-i k \tau) e^{i k \tau} ~, \tau_c < \tau < \tau_d ~,\\
    \calR_{k}(\tau)&=\alpha_{k} \frac{H}{\sqrt{4 \epsilon k^{3}}}(1+i k \tau) e^{-i k \tau}+\beta_{k} \frac{H}{\sqrt{4 \epsilon k^{3}}}(1-i k \tau) e^{i k \tau}~, \tau > \tau_d ~,
\end{align}

and their derivatives
\begin{equation}
    \begin{aligned}
        \calR_{k}^{\prime}(\tau)=& \alpha^{r}_{k} \left[\frac{H}{\sqrt{4 \epsilon k^{3}}} k^{2} \tau e^{-i k \tau}+\frac{\eta}{2 \tau} \frac{H}{\sqrt{4 \epsilon k^{3}}}(1+i k \tau) e^{-i k \tau}\right] \\
        &+\beta^{r}_{k}\left[\frac{H}{\sqrt{4 \epsilon k^{3}}} k^{2} \tau e^{i k \tau}+\frac{\eta}{2 \tau} \frac{H}{\sqrt{4 \epsilon k^{3}}}(1-i k \tau) e^{i k \tau}\right] ~, \tau_c < \tau < \tau_d
    \end{aligned}
\end{equation}
\begin{equation}
    \begin{aligned}
        \calR_{k}^{\prime}(\tau)=& \alpha_{k}\left[\frac{H}{\sqrt{4 \epsilon k^{3}}} k^{2} \tau e^{-i k \tau}+\frac{\eta}{2 \tau} \frac{H}{\sqrt{4 \epsilon k^{3}}}(1+i k \tau) e^{-i k \tau}\right] \\
        &+\beta_{k}\left[\frac{H}{\sqrt{4 \epsilon k^{3}}} k^{2} \tau e^{i k \tau}+\frac{\eta}{2 \tau} \frac{H}{\sqrt{4 \epsilon k^{3}}}(1-i k \tau) e^{i k \tau}\right] ~, \tau > \tau_d ~,
    \end{aligned}
\end{equation}
where $\alpha^{r}_{k}$, $\alpha_{k}$, $\beta^{r}_{k}$ and $\beta_{k}$ are given by Eq.~\eqref{C Match} and \eqref{D Match}. It is now easy to check that our results satisfy the well-known gauge transformation relation $\calR = (H/\dot\phi)\delta\phi$. After taking the $\epsilon_r\rightarrow +\infty$ limit, we can directly get the long wavelength behaviour of the mode function after $\tau_c$
\begin{equation}
    \calR_k(\tau) \simeq - \frac{H}{\sqrt{4 \epsilon_V k^3}}\frac{ \left(g^2 h-6\right)}{6 g } ~,\quad k < k_c ~.
\end{equation}
Meanwhile, we can also get the short-wavelength behaviour
\begin{equation}
    \calR_k(\tau) = \frac{H}{\sqrt{4\epsilon_V k^3}}\frac{\left(g^2+1\right)}{2g} - \frac{H}{\sqrt{4\epsilon_V k^3}}\frac{\left(g^2-1\right) e^{-2 i k \tau_c}}{2 g} ~, \quad k > k_c ~.
\end{equation}

\subsection{The scaling behaviour when including an inflection points}\label{Inflection point and scaling}

In this appendix, we include the discussions about the much realistic model with an inflection point as mentioned at the beginning of Section \ref{Inflection point}. Following the same mode function matching procedure, here we choose another characteristic comoving wavenumber $k_s$ which crosses the horizon at $\phi=\phi_s$, i.e. $\tau=\tau_s$. Let us focus on the long wavelength modes $k \lesssim k_s$, and then the Bogolyubov coefficients can be written as
\begin{equation}\label{Inflection matching}
    \alpha_k+\beta_k \simeq -\frac{gh}{6} \(\frac{k_s}{k_c}\)^3 - \frac{2}{5g}\Big(1+O\(g^2|h|,(k_s/k_c)^2\)\Big) \(\frac{k}{k_s}\)^2 + O\((k/k_s)^3\) ~,
\end{equation}
If we further require $k \ll k_s$, which means that we focus on the extremely large scale modes, then we can successfully recover the standard power spectrum from single-field SR inflation
\begin{equation}
    P_{\mathcal{R}_{\text{G}}}(k) \simeq \frac{H^2}{8\pi^2\epsilon_S} ~,\quad k \ll k_s ~,
\end{equation}
where $\epsilon_S$ is the slow-roll parameter during the first SR stage in our model and $\mathcal{R}_{\text{G}}$ represents the Gaussian part curvature perturbation. Meanwhile, in the $g^2|h| \ll 1$ region, where curvature perturbation is dominated by the effects of the step, Eq.~\eqref{Inflection matching} also tells us the scaling behaviour, i.e. the spectrum index $n_s$ when $k \lesssim k_s$
\begin{equation}
    n_s -1 = \frac{d \log P_{\mathcal{R}_{\text{G}}}(k)}{d \log k} \simeq 4 ~, \quad k \lesssim k_s ~.
\end{equation}
This is in agreement with the arguments from \cite{Byrnes:2018txb} that the typical possible growth in such a transition model has a spectrum index $n_s-1=4$. A more general discussion of the steepest growth of power spectrum in a single-field inflation scenario can be found in \cite{Carrilho:2019oqg,Ozsoy:2019lyy}.

For the short wavelength modes $k \gg k_s$, the perturbations oscillate following the frequency $\omega^2 \sim k^2$, and thus the inflection point can be ignored.



\bibliographystyle{JHEP}
\bibliography{references}
\end{document}